\documentclass{aa}  
\usepackage{array}
\usepackage{graphicx}
\usepackage{multirow}
\usepackage{color}
\usepackage{xcolor}
\usepackage{hyperref}
\usepackage{capt-of}

\usepackage[capitalize]{cleveref}
\usepackage{txfonts}
\renewcommand{\arraystretch}{1.2}

\def\N1316{NGC\,1316}
\def\N1404{NGC\,1404}

\def\4U{4U~1735$-$444}

\def\aap{{A\&A}}

\def\apjl{{ApJ}}

\def\apjs{{ApJS}}

\def\arcsec{\ifmmode '' \else $''$\fi}

\def\arcsecpoint{\ifmmode ''\!. \else $''\!.$\fi}

\def\kms{\ifmmode {\rm km\ s}^{-1} \else km s$^{-1}$\fi}
\def\Msun{\ifmmode {\rm M}_{\odot} \else M$_{\odot}$\fi}
\def\Lsun{\ifmmode {\rm L}_{\odot} \else L$_{\odot}$\fi}
\def\Zsun{\ifmmode {\rm Z}_{\odot} \else Z$_{\odot}$\fi}

\def\ergscm2{ergs\,s$^{-1}$\,cm$^{-2}$}
\def\icm3{{\rm cm}^{-3}}
\def\icm2{{\rm cm}^{-2}}
\def\qo{\ifmmode q_{\rm o} \else $q_{\rm o}$\fi}
\def\Ho{\ifmmode H_{\rm o} \else $H_{\rm o}$\fi}
\def\ho{\ifmmode h_{\rm o} \else $h_{\rm o}$\fi}

\def\vFWHM{\ifmmode v_{\mbox{\tiny FWHM}} \else
            $v_{\mbox{\tiny FWHM}}$\fi}
\def\CCF{\ifmmode F_{\it CCF} \else $F_{\it CCF}$\fi}
\def\ACF{\ifmmode F_{\it ACF} \else $F_{\it ACF}$\fi}
\def\Halpha{\ifmmode {\rm H}\alpha \else H$\alpha$\fi}
\def\Hbeta{\ifmmode {\rm H}\beta \else H$\beta$\fi}
\def\Hgamma{\ifmmode {\rm H}\gamma \else H$\gamma$\fi}
\def\Hdelta{\ifmmode {\rm H}\delta \else H$\delta$\fi}
\def\Lya{\ifmmode {\rm Ly}\alpha \else Ly$\alpha$\fi}
\def\Lyb{\ifmmode {\rm Ly}\beta \else Ly$\beta$\fi}
\def\Lyg{\ifmmode {\rm Ly}\beta \else Ly$\gamma$\fi}

\def\ciii{\ifmmode {\rm C}\,{\sc iii} \else C\,{\sc iii}\fi}
\def\civ{\ifmmode {\rm C}\,{\sc iv} \else C\,{\sc iv}\fi}
\def\cv{\ifmmode {\rm C}\,{\sc v} \else C\,{\sc v}\fi}
\def\cvi{\ifmmode {\rm C}\,{\sc vi} \else C\,{\sc vi}\fi}

\def\nvii{N\,{\sc vii}}
\def\nvivii{N\,{\sc vi-vii}}

\def\o5007{[O\,{\sc iii}]\,$\lambda5007$}

\def\ovii{O\,{\sc vii}}
\def\oviii{O\,{\sc viii}}
\def\oviiviii{O\,{\sc vii-viii}}

\def\neix{Ne\,{\sc ix}}
\def\nex{Ne\,{\sc x}}
\def\neixx{Ne\,{\sc ix-x}}

\def\mgxi{Mg\,{\sc xi}}
\def\mgxii{Mg\,{\sc xii}}

\def\fexvii{Fe\,{\sc xvii}}

\def\fexxii-iii{Fe\,{\sc xxii-xxiii}}

\usepackage{hyperref}
\begin{document} 

   \title{XMM-Newton multi-year campaign on NGC 55 ULX-1}

   \subtitle{Resolving the wind and its variability with RGS}

   \titlerunning{NGC 55 ULX-1 RGS}

   \author{C. Pinto
          \inst{1}\fnmsep\thanks{ciro.pinto@inaf.it}
          \and
          S. Caserta\inst{1,2}
          \and
          F. Barra\inst{1,2,3}
          \and
          Y. Xu\inst{1,4}
          \and
          D. Barret\inst{5}
          \and
          P. Kosec\inst{3}
          \and
          N. La Palombara\inst{6}
          \and
          A. Marino\inst{7}
          \and \\
          F. Pintore\inst{1}
          \and
          A. Riggio\inst{8}
          \and
          T. P. Roberts\inst{9}
          \and
          C. Salvaggio\inst{10}
          \and
          L. Sidoli\inst{6}
          \and
          R. Soria\inst{11}
          \and
          D. J. Walton\inst{12}
          }

      \institute{INAF -- IASF Palermo, Via U. La Malfa 153, I-90146 Palermo, Italy
        \and
            Universit\`a degli Studi di Palermo, Dipartimento di Fisica e Chimica, 
            via Archirafi 36, I-90123 Palermo, Italy
        \and
            Center for Astrophysics -- Harvard \& Smithsonian, Cambridge, MA, USA
        \and 
             Department of Astronomy \& Physics, Saint Mary's University, 923 Robie Street, Halifax, NS B3H 3C3, Canada
        \and
            Université  de Toulouse, CNRS, IRAP, 9 Avenue du colonel Roche, BP 44346, F-31028 Toulouse Cedex 4, France
        \and
            INAF -- Istituto di Astrofisica Spaziale e Fisica Cosmica di Milano, Via A. Corti 12, 20133 Milano, Italy
        \and 
            Institute of Space Sciences (ICE), CSIC, Campus UAB, Carrer de Can Magrans s/n, E-08193, Barcelona, Spain
        \and 
            INAF -- Osservatorio Astronomico di Cagliari, localita Poggio dei Pini, Strada 54, 09012 Capoterra, Italy
        \and 
            Department of Physics, Centre for Extragalactic Astronomy, Durham University, South Road, Durham DH1 3LE, UK
        \and 
            INAF -- Osservatorio Astronomico di Brera, Via E. Bianchi 46, 23807 Merate, (LC), Italy
        \and 
            INAF -- Osservatorio Astrofisico di Torino, Strada Osservatorio 20, I-10025 Pino Torinese, Italy
        \and
            Centre for Astrophysics Research, University of Hertfordshire, College Lane, Hatfield AL10 9AB, UK
             }

   \date{Received July 28, 2025; accepted December 16, 2025}
 
  \abstract
   {Winds are an important ingredient in the evolution of X-ray binary (XRB) systems, particularly those at high accretion rates such as ultra-luminous X-ray sources (ULXs), because they may regulate the accretion of matter onto the compact object.}
   {We aim at understanding the properties of ULX winds and their link with the source spectral and temporal behavior.}
   {We performed high-resolution X-ray spectroscopy of the variable source NGC 55 ULX-1 to resolve emission and absorption lines as observed with XMM-Newton at different epochs. Optically-thin plasma models are used to characterise the wind.}
   {We confirmed and thoroughly strengthened previous evidence of outflows in NGC 55 ULX-1. The presence of {radiative recombination} signatures and the ratios between the fluxes of the emission lines favours photoionisation balance and low-to-moderate densities, which confirm that the lines originate from classical XRB disc winds. An in-depth parameter space exploration shows line emission from a slowly moving, cool, and variable plasma perhaps associated with a thermal wind. Mildly-relativistic Doppler shifts {(about $-0.15c$)} associated with the absorption lines confirm, at higher confidence, the presence of powerful, radiatively-driven, winds.
}
   {The comparison between results obtained at different epochs revealed that the wind responds to the variability of the underlying continuum and these variations may be used to understand the actual accretion regime and the nature of the source.}


   \maketitle

\section{Introduction}\label{sec:intro}

Ultra-luminous X-ray sources (ULXs) occupy the brighest-end portion of the luminosity function of X-ray binaries (for recent reviews on ULXs, see \citealt{King2023,Pinto2023a}). Shining in the X-ray band (0.3--10 keV) above $10^{39} \rm erg \ s^{-1}$, i.e. the Eddington limit for a stellar-mass black hole, they have been a matter of lively debate in the last 2 decades, particularly thanks to important discoveries enabled with the X-ray telescopes onboard the XMM-Newton, Chandra, NUSTAR and Swift satellites. Nowadays we know that most ULXs in the nearby Universe are powered by stellar-mass compact objects, such as neutron stars (NS) and black holes (BH), accreting above their Eddington limit and, therefore, represent the most extreme mass-transfer phase of interacting X-ray binaries (XRBs). This was particularly cemented after the discovery of coherent pulsations in a handful of nearby bright sources (dubbed PULXs; see e.g. \citealt{Bachetti2014,Fuerst2016,Israel2017a}). 

High-quality broadband spectra of ULXs often peak in the X-ray band and show a peculiar curvature below 10 keV, which favoured a scenario of high mass transfer, even before the discovery of the pulsations (see, e.g., \citealt{Gladstone2009,Bachetti2013,Walton2014}). These spectra also revealed the presence of unresolved features in the soft X-ray band and, particularly, around 1 keV (see, e.g., \citealt{Soria2004,Stobbart2006,Middleton2015b}) which have then been unambiguously identified as a forest of emission and absorption lines with the high-resolution grating spectrometers onboard XMM-Newton and Chandra (see, e.g., \citealt{Pinto2016nature,vdEijnden2019,Kosec2021}). These unveiled the long-sought winds driven by strong radiation pressure which were the ultimate proof of supercritical accretion (see, e.g., \citealt{Ohsuga2005,Poutanen2007}). 

ULX winds have Doppler motions reaching velocities of about $0.2c$ and have often been invoked as the engine powering the interstellar bubbles found around many ULXs (see, e.g., \citealt{Pakull2010,Gurpide2022}). Moreover, super-Eddington winds are expected to be equatorial and highly clumpy and are likely able to scatter the hard X-ray photons coming from the innermost regions of the accretion disk, thereby influencing the way the ULX appears depending on the source view angle (see, e.g., \citealt{Middleton2015a,Kobayashi2018}). For this reason, multi-epoch observations of variable and bright ULXs provide an ideal workbench to study the geometry and structure of super-Eddington winds, their influence on the source appearance, and overall kinetic budget.

\section{NGC 55 multi-year monitoring}\label{sec:ngc55ulx}

NGC 55 ULX-1 is the brightest X-ray source in its host galaxy. At a distance of about 1.94 Mpc\footnote{https://ned.ipac.caltech.edu/}, this source has a peak X-ray luminosity (0.3--10 keV) of about $4 \times 10^{39} \ \rm erg \ s^{-1}$ (see, e.g., \citealt{Gurpide2021}). The long-term Swift / XRT monitoring performed in 2013 showed substantial variability over a few weeks: a factor 3--4 in the 0.3--10 keV band with the spectrum appearing harder when brighter, as already shown in many other ULXs, which has been ascribed either to having a clearer view of the inner accretion flow (i.e. viewing this through a less dense wind) when the flux is higher, or to changes in the accretion rate through the disc \citep{Pintore2015}. On short ($\sim100$\,s) time scales, Chandra and XMM-Newton light curves taken during the highest-flux epochs have shown sharp drops, during which the source exhibits softer spectra and significantly suppressed flux above $\sim$2 keV (\citealt{Stobbart2006}). This behavior is more often seen in supersoft ULX sources (SSUL or ULS) and is likely due to wind clumps that temporarily obscure the innermost part of the disc \citep{Urquhart2016}. For this reason, NGC 55 ULX-1 has been considered to be at an inclination slightly lower than SSUL (e.g.,  \citealt{Pinto2017}, hereafter P17).
Due to its flux and spectral variability, NGC 55 ULX-1 is an ideal target for studying the behaviour of the wind during different epochs and its role in the X-ray spectral appearance. Besides, its X-ray luminosity oscillating around $2 \times 10^{39} \ \rm erg \ s^{-1}$ means that we may use it as a proxy for a compact object behavior crossing the Eddington limit, in the case the accretor is a stellar-mass BH as indicated by recent results \citep[][hereafter B22]{Barra2022}.

\subsection{Swift / XRT long-term monitoring}
\label{sec:swift}

   \begin{figure*}[h]
   \centering
   \includegraphics[scale=0.5, angle=0, width=18cm]{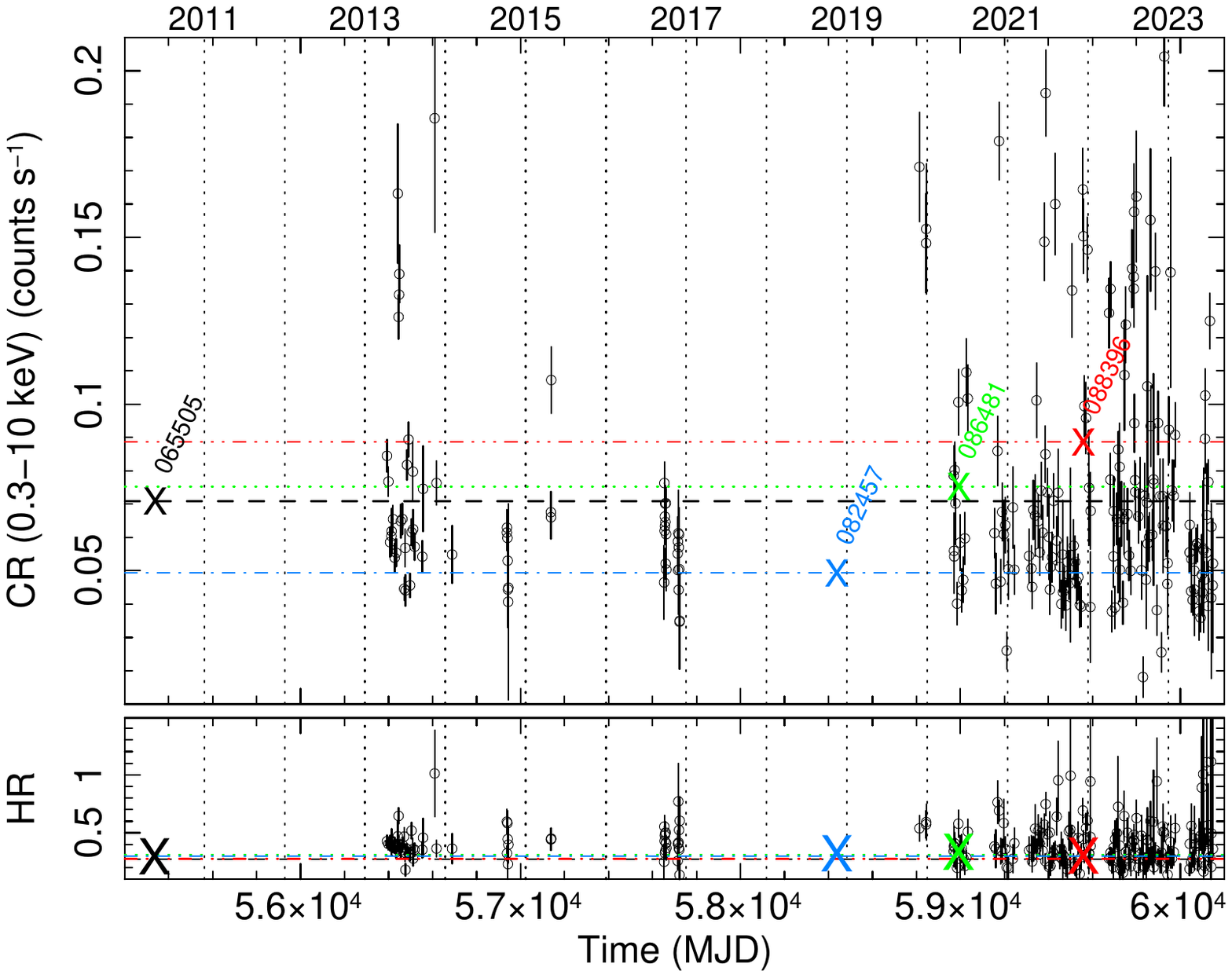}
   \vspace{-0.1cm}
   \caption{{Swift/XRT long-term light curve (top panel) and hardness-ratio curve of NGC 55 ULX-1 (bottom panel). The hardness ratio is computed as the ratio between the counts in 2--10 keV / 0.3--2 keV energy bands. The values for the 4 on-axis deep XMM-Newton observations are shown with horizontal lines and the times of the events with "X" marks followed by the corresponding obsid}.}
              \label{fig:swift}
   \vspace{-0.1cm}
    \end{figure*}

In order to study NGC 55 ULX-1 with deep (about 1 orbit) XMM-Newton observations in different flux regimes and to understand its long-term behaviour, we regularly monitored the source with the Swift XRT (\citealt{Gehrels2004}) from early 2020 to the end of 2023. The XRT light curve was retrieved through the web tool\,\footnote{https://www.swift.ac.uk/user$\_$objects/} \citep{Evans2009}. {Two long-term light curves were also extracted in the 2--10 keV and 0.3-2 keV bands to compute a hardness curve (hard / soft ratio), see Fig.\,\ref{fig:swift}}. The band split is chosen at 2\,keV because most variability is seen below this energy (see, e.g., B22). Variability of up to an order of magnitude can be seen in the count rate on timescales of a few days.

NGC 55 ULX-1 exhibits an absorbed 0.3--10\,keV flux ranging from $1\times 10^{-12}$ erg s$^{-1}$ cm$^{-2}$ during the dips to above $6 \times 10^{-12}$ erg s$^{-1}$ cm$^{-2}$ during flaring events (P17). A long archival (2010) observation with XMM-Newton caught NGC 55 ULX-1 at an intermediate flux corresponding to an XRT count rate of 0.07\,c/s (converted via WEBPIMMS\,\footnote{https://heasarc.gsfc.nasa.gov/cgi-bin/Tools/w3pimms/w3pimms.pl} through an absorbed powerlaw model, e.g. \citealt{Pintore2015}). Our first new XMM-Newton visit observed the source during a low-flux period in November 2018 and we then tried to trigger two more observations during high-flux epochs, catching it at an intermediate flux in May 2020 (comparable to the archival 2010 data) and in the tail of a flaring period in December 2021. 

\subsection{First results of the XMM-Newton campaign}
\label{sec:first_results_of_the_campaign}

Our new observations enabled us to investigate the broadband properties of NGC 55 ULX-1 and to find strong evidence for a second, transient, ultra-luminous X-ray sources (dubbed NGC 55 ULX-2) in the same galaxy. Regarding the new transient ULX, in \citet{Robba2022} we reported that the object is closer to the galactic centre and has a bolometric luminosity peak of about $3 \times 10^{39} \ \rm erg \ s^{-1}$ in multiple epochs. Its spectral shape and variability fits within the subclass of soft ultra-luminous X-ray sources (SUL). In B22, instead, we reported a detailed spectral modelling of ULX-1 with the high-statistics XMM-Newton / EPIC spectra. A double thermal model showed luminosity-temperature (L-T) trends in broad agreement with the L$ \ \propto$ T$^{4}$ relationship expected from a thin-disc model. However, small but significant deviations at the highest luminosities suggest that the disc structure is changing e.g. due to an expansion or the contribution from the wind at accretion rates higher than the Eddington (${\dot M_{Edd}}$) or even supercritical (${\sim 2-3 \dot M_{Edd}}$), which in turn would indicate a stellar-mass black hole as the accretor.

In this paper, we focus on the properties of ULX-1 winds by performing a detailed high-resolution X-ray spectroscopy with the reflection grating spectrometers (RGS) onboard XMM-Newton. We include the archival data (2010) already shown in P17 in order to cross-check and validate our results. The rest of the manuscript is organised as follows. We report the XMM-Newton data reduction in Sect.\,\ref{sec:data_reduction} and the spectral modelling in Sect.\,\ref{sec:data_modelling}. We discuss our results in Sect.\,\ref{sec:discussion} and provide our conclusions in Sect.\,\ref{sec:conclusions}. All uncertainties are given at 1$\sigma$ (68\,\% level).

\section{Data reduction}\label{sec:data_reduction}

\begin{figure}
   \centering
   \includegraphics[scale=0.35]{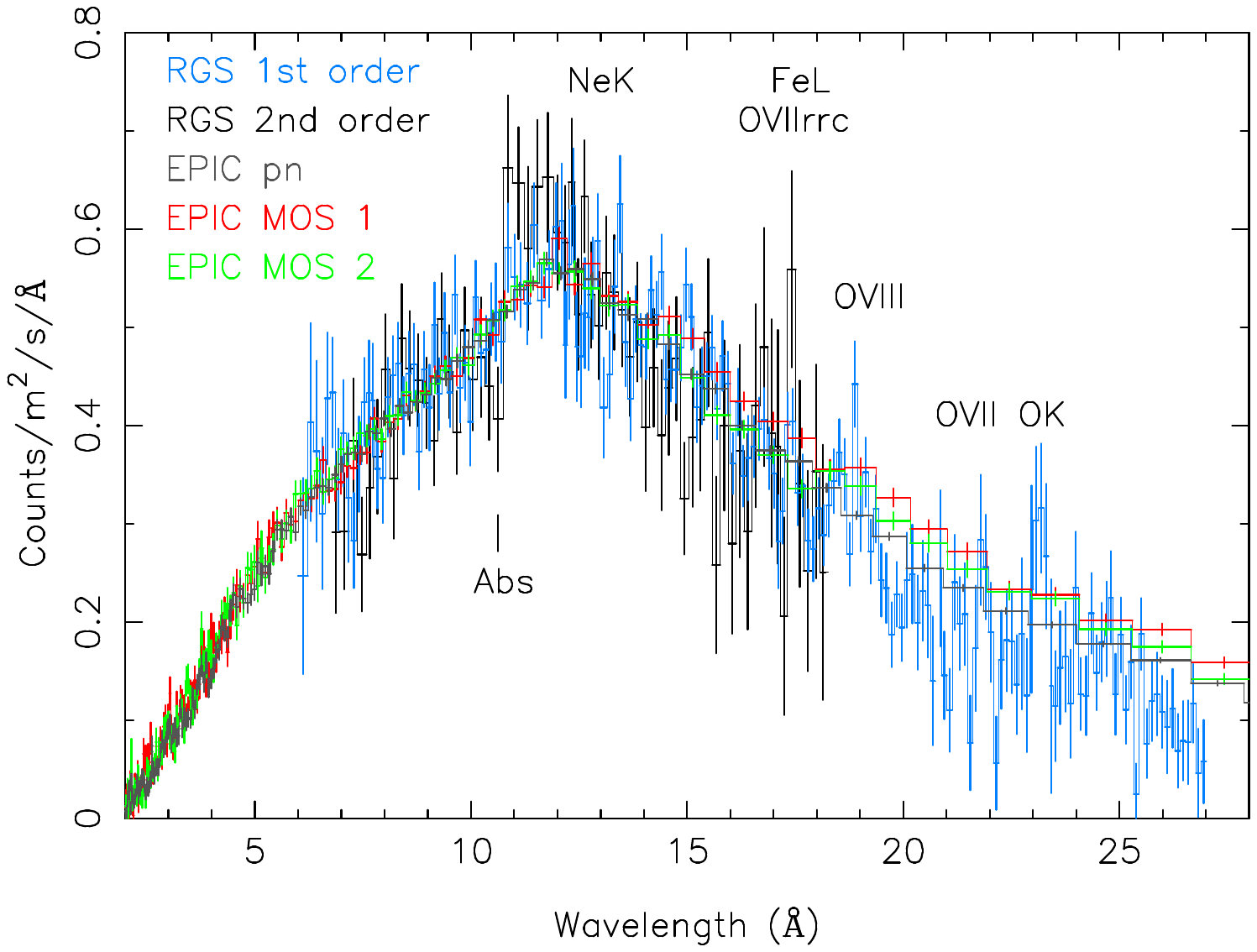}
   \vspace{-0.5cm}
      \caption{XMM-Newton combined spectra of ULX-1 obtained by stacking those from the four deep, on-axis, observations: RGS $1^{\rm st}$ (light-blue) and $2^{\rm nd}$-order (black), EPIC-pn (grey), MOS\,1 (red) and MOS\,2 (green).}
         \label{fig:spec_tavg}
   \vspace{-0.3cm}
   \end{figure}

In this work, we used the four on-axis, deep, XMM-Newton observations of NGC 55 ULX-1. These enable us to obtain individual high-resolution RGS spectra along with the high signal-to-noise EPIC spectra, which are necessary to find and identify wind features, and track down any changes at different epochs. The observations IDs are 065505(0101) (05/2010), 082457(0101) (11/2018), 086481(0101) (05/2020) and 088396(0101) (12/2021). 
To avoid redundancy, we do not show all details regarding the EPIC data reduction, since it was already reported in B22. Briefly, raw data were retrieved from the XMM-Newton Science Archive (XSA)\,\footnote{https://www.cosmos.esa.int/web/XMM-Newton/xsa} and reduced with the \textit{Science Analysis System} ({\scriptsize{SAS}}) version 22.1.0\,\footnote{https://www.cosmos.esa.int/web/XMM-Newton} with recent calibration files (February 2025). EPIC MOS and pn data were processed with the {\scriptsize{EPPROC}} and {\scriptsize{EMPROC}} tasks according to standard procedures regarding patterns and Solar flares removal. The clean exposure times were identical to those reported in B22. The source spectra were extracted using a circular region of 20'' 
radius (to keep background contamination low) centred on the X-ray position estimated with Chandra  (RA: 00$^{h}$ 15$^{m}$ 28.89$^{s}$ DEC: -39$^{\circ}$ 13' 18.8'', \citealt{Gladstone2009}) while the background was selected from a larger circular region a few arc minutes away from the source and from the copper ring in the same chip. The response matrices and effective area files were extracted with the {\scriptsize{RMFGEN}} and {\scriptsize{ARFGEN}} tasks.
We also stacked the EPIC-pn, MOS\,1 and MOS\,2 spectra from the four observations using {\scriptsize{EPICSPECCOMBINE}} in order to obtain 3 time-averaged spectra with high signal-to-noise ratio for the SUL regime of NGC 55 ULX-1 (see Fig.\,\ref{fig:spec_tavg}).
 
The RGS data reduction was performed with the {\scriptsize{RGSPROC}} which also extracts spectra and response/area files. We extracted the $1^{\rm st}$ and $2^{\rm nd}$-order RGS spectra in a cross-dispersion region of 0.8' width, centred on the source coordinates and the background spectra by selecting photons
beyond the 98\% of the source point spread function. The background regions do not overlap with bright sources. We filtered out periods affected by contamination from Solar flares by selecting background-quiescent intervals in the light curves of the RGS 1,2 CCD\,9
({\it i.e.}, $\gtrsim1.7$\,keV) with a standard count rate below 0.2 c/s. As is typically found, Solar flares affected the RGS data to a much lower extent than the EPIC data. The total clean exposure times for each RGS spectrometer are 120, 135, 129 and 122\,ks for obs-id 0655, 0824, 0864 and 0883, respectively. The four observations yield RGS\,1+2 $1^{\rm st}$-order spectra with a total of 5773, 4545, 7853 and 5465 source-net counts, respectively. Typically, $\gtrsim5000$ counts are necessary to detect the strongest lines \citep{Kosec2021}. 

\begin{figure*}
   \centering
   \includegraphics[scale=0.55, width=17.cm, height=10cm]{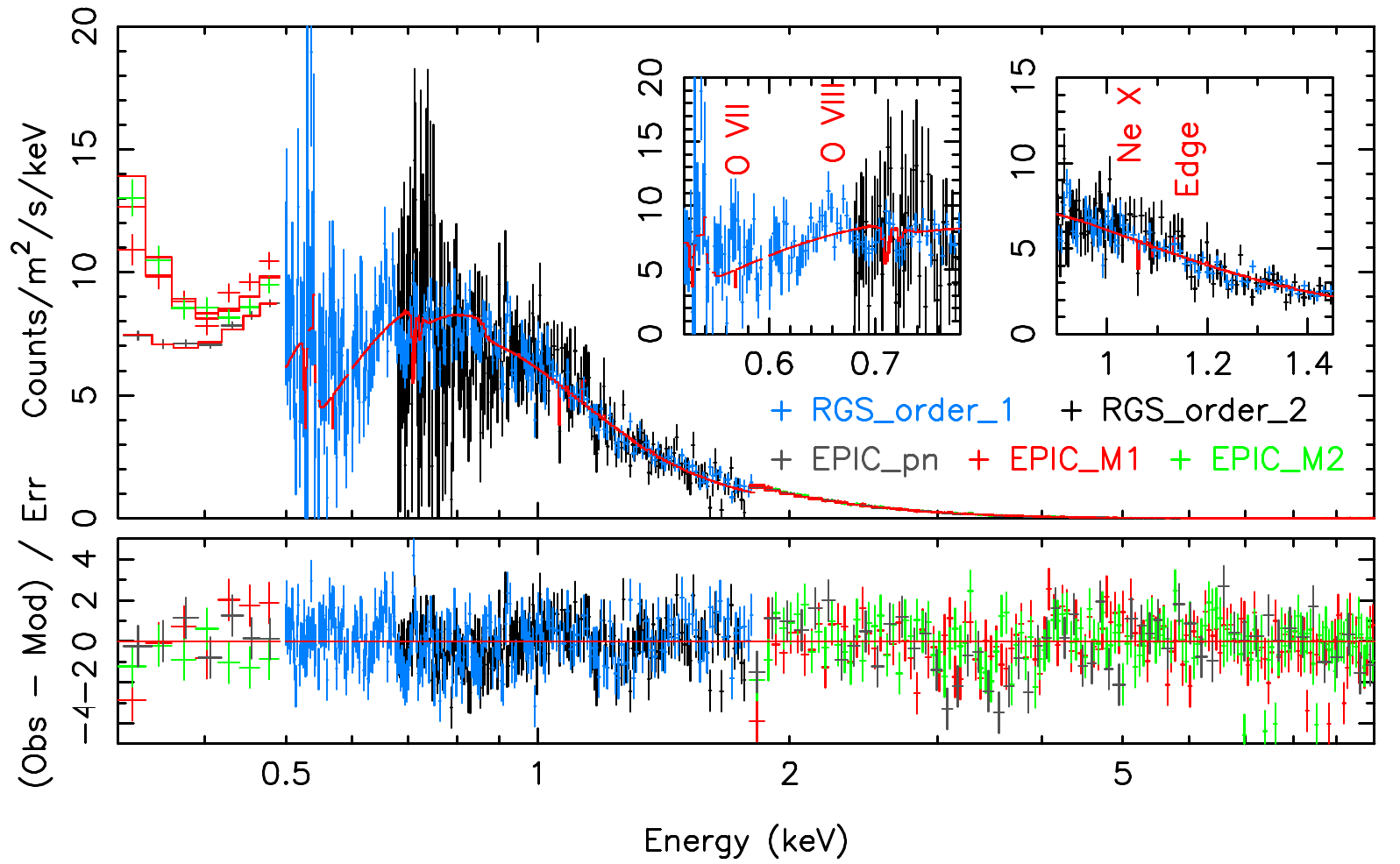}
   \vspace{-0.1cm}
      \caption{XMM-Newton time-averaged (on-axis) spectra and continuum model. Labels are same as in Fig.\,\ref{fig:spec_tavg}. The EPIC data were ignored between 0.5-1.8 keV in order to fully employ the high spectral resolution of RGS and decrease degeneracy between models of line emission / absorption.}
         \label{fig:spec_tavg_fit}
   \vspace{-0.1cm}
   \end{figure*}

Through the {\scriptsize{RGSCOMBINE}} task we also produced a combined spectrum for the $1^{\rm st}$-order, which yields a total of 23636 counts and an exposure of 506\,ks, which is one of the 10 deepest ULX RGS spectra and the $2^{\rm nd}$ deepest SUL spectrum after that of NGC\,5408 ULX-1 \citep{Pinto2016nature,Kosec2021,Pinto2023a}. Finally, we also combined all the $2^{\rm nd}$-order RGS\,1 and 2 spectra obtaining a 7536 counts spectrum with good statistics, which provides the clearest view around 1 keV (or 12\,{\AA}) with a resolving power $\lambda / \Delta \lambda > 300$ and increasing at lower energies; see Fig.\,\ref{fig:spec_tavg} and \ref{fig:spec_tavg_fit}.
 
{We note that the spectral stacking is necessary to speed-up computation, particularly to run Monte Carlo (MC) simulations that are suitable for estimating the signification of the line detection. This already takes about a month for an accurate scanning of spectral lines in a statistical sample of simulated spectra (see Sect.\,\ref{sec:monte_carlo_simulations} and the dedicated discussion in \citealt{Pinto2023a}).}

{We also had a quick look at the data from the Optical Monitor (OM) onboard XMM-Newton to search for an optical/UV counterpart of ULX-1. The source was undetected in the images provided by all filters (V, UVW1, UM2, and UVW2) and by stacking those from the same filters. This was expected given its very low flux in observations with Hubble Space Telescope (HST magnitude $\gtrsim23$; \citealt{Gladstone2013}, \citealt{Zhou2023}).}

\section{Spectral modelling}\label{sec:data_modelling}

The spectra were modelled with the {\scriptsize{SPEX}} fitting package 3.08.01 \citep{kaastraspex}, which has a full suite of line-emitting / absorbing optically-thin plasma. The spectra were grouped according to the optimal binning directly in {\scriptsize{SPEX}}, {which provides at least 1 count per bin and bin-size $\geq 1/3$ of the spectral resolution,} and fit by minimising the $C-$statistics \citep{Cash1979,Kaastra2017}.
All emission components were corrected for absorption from the circumstellar and interstellar medium with the \texttt{hot} model (freezing the gas temperature to $10^{-6}$ keV, which provides a neutral gas in {\scriptsize{SPEX}}). To limit the computation time, we adopted the recommended Solar abundances of \citet{Lodders2009}, which are default in {\scriptsize{SPEX}}, for all emitting and absorbing plasma components. This is generally a good approximation, although might have an impact on the detection of the winds given that their abundances may differ from the Solar pattern as e.g. recently shown in Holmberg II (UL)X-1 \citep{Barra2024}.
Each model is fitted simultaneously to the EPIC MOS 1, 2 / pn and RGS $1^{\rm st} / 2^{\rm nd}$ order spectra with a free multiplicative constant that accounts for the typical 1-5\,\% cross-calibration uncertainties. 

Both the combined spectra and those of the individual observations were analysed with the following scheme: continuum modelling, gaussian line scan, physical model scan, and best-fit physical model. We first focussed on the high-statistics, time-averaged, stacked spectrum in order to identify the main features / lines in the RGS spectrum and search for evidence of outflows. Afterwards, we performed an analysis of the individual observations to search for any variability in the lines and outflow parameters (see Sect. \ref{sec:individual_obs}).

\subsection{Time-averaged spectrum}\label{sec:tavg}

The combined EPIC MOS 1, 2 / pn and RGS $1^{\rm st} / 2^{\rm nd}$ spectra are shown in Fig.\,\ref{fig:spec_tavg}. 
By visual inspection it is easy to spot the well-known {\oviii} H-like emission line around 19\,{\AA}, the He-like {\ovii} emission feature near 22\,{\AA}, and the 23\,{\AA} oxygen K-edge which were previously found in the archival data and in many more ULXs (P17, \citealt{Kosec2021}). Due to its higher resolution, the $2^{\rm nd}$ order spectrum resolves the feature around 1 keV in what looks like a sharp drop below 11\,{\AA}. This is similar to that observed at lower resolution in SSUL spectra and previously interpreted as an edge due to photon absorption in the wind (see, e.g., \citealt{Urquhart2016,Pinto2021}).

\subsubsection{Continuum modelling and spectral energy distribution}\label{sec:continuum_and_SED}

RGS and EPIC time-averaged spectra were simultaneously fitted. In order to fully employ the high resolution of RGS and decrease the degeneracy between models, we ignored the EPIC data within 0.5-1.8 keV where the RGS source spectrum is well above the background. A two-component emission model was adopted following P17, i.e., a cool 0.16 $\pm$ 0.01 keV blackbody (\texttt{bb}) and a warmer 0.70 $\pm$ 0.01 keV blackbody modified by coherent Compton scattering (\texttt{mbb}) for the broader/harder component produced by the hotter super-Eddington inner disc. The best-fit column density is (2.5 $\pm$ 0.1) $\times \, 10^{21}$ cm$^{-2}$. This time-averaged value was also adopted later for the fit of the individual observations, as it is not expected to vary significantly (P17, B22). The 0.3-10 keV (unabsorbed) luminosities of the cool and warm thermal components are around $9.4 \times 10^{38}$ erg s$^{-1}$ and $5.6 \times 10^{38}$ erg s$^{-1}$ for a total of $1.5 \times 10^{39}$ erg s$^{-1}$ while the total observed flux is around $1.5 \times 10^{-12}$ erg s$^{-1}$ cm$^{-2}$. The best-fit continuum model is shown in Fig.\,\ref{fig:spec_tavg_fit}. Sharp narrow residuals can be seen along with the drop above 1 keV mentioned above.

Later, in Sect.\,\ref{sec:model_scan} we show the application of photoionisation (and collisional) models. For the computation of photoionisation equilibrium, a broadband spectral energy distribution (SED) is necessary. Following \cite{Pinto2021} approach on a similar source (soft / supersoft NGC 247 ULX-1), we extrapolated the (unabsorbed) best-fit double thermal model down to $10^{-4}$ keV and up to $10^{2}$ keV, for a total, bolometric luminosity of $1.7 \times 10^{39}$ erg s$^{-1}$. 
Most of the flux for this type of object is anyway yielded in the 0.3-10 keV range, which results in low systematic uncertainties due to the broadband extrapolation (see e.g. \citealt{Pinto2020a}).

\subsubsection{Gaussian line scan}\label{sec:gaussian_scan}

In order to search for narrow spectral lines, we performed a gaussian line scan consisting of a fit with a moving gaussian line with the method outlined in (\citealt{Pinto2016nature}, P17). On top of the \texttt{bb}+\texttt{mbb} continuum model, we added a gaussian with the centroid shifting from 0.5 to 7 keV (above which EPIC spectra are background-dominated), exploring the presence of lines also in the EPIC spectra, particularly above 2-3 keV where the CCD spectral resolution progressively increases. We used a logarithmic grid with {500 (3000) steps for the adopted velocity dispersion ($\sigma_{\rm V} = \rm FWHM / 2.355$) of 5000 (100) km/s, respectively. This was used to mediate between the resolving power needed for a given line width and the computation time as well as to search for lines from plasma going from thermal (e.g. broadening around 100 km/s) to more extreme outflows as seen in other ULXs, particularly for a (S)SUL regime similar to NGC 55 ULX-1 (e.g. NGC 247 ULX-1, \citealt{Pinto2021}).}

   \begin{figure}
   \centering
   \includegraphics[scale=0.725]{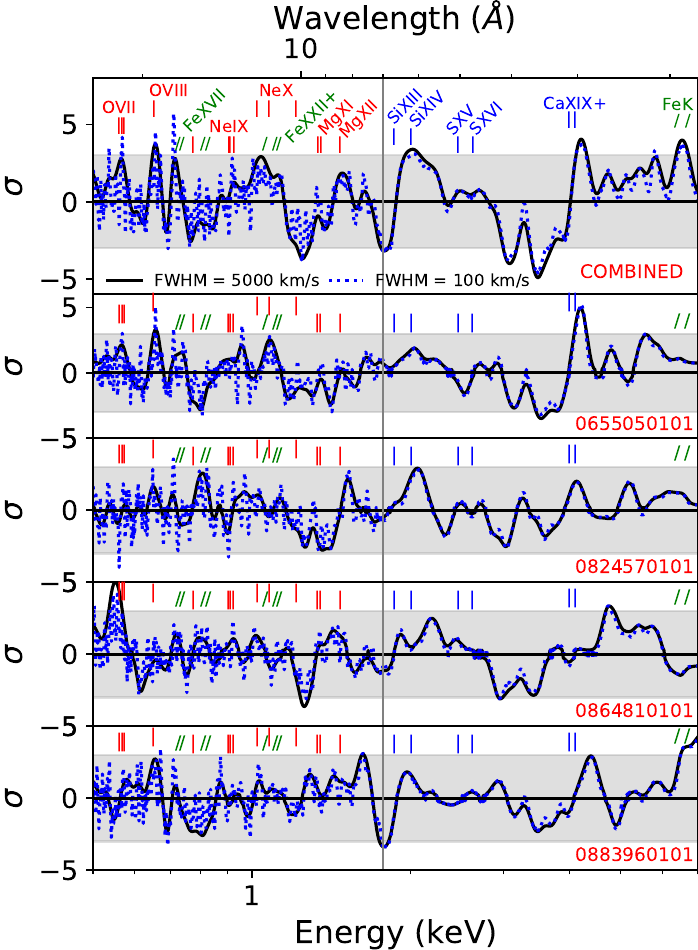}
   \vspace{-0.1cm}
      \caption{Gaussian line scan of the XMM-Newton time-averaged spectrum (top panel) and spectra of the individual observations. Labelled are the energy centroids of some among the most common X-ray spectral lines {with Fe K referring to the Fe I fluorescence and the {Fe\,{\sc xxv}} resonant lines. The grey vertical line indicates the separation between RGS and EPIC}.
      The grey-shaded areas indicate the $3\,\sigma$ single-trial significance.}
         \label{fig:spec_all_gaus}
   \vspace{-0.2cm}
   \end{figure}

   \begin{figure}
   \centering
   \includegraphics[scale=0.33]{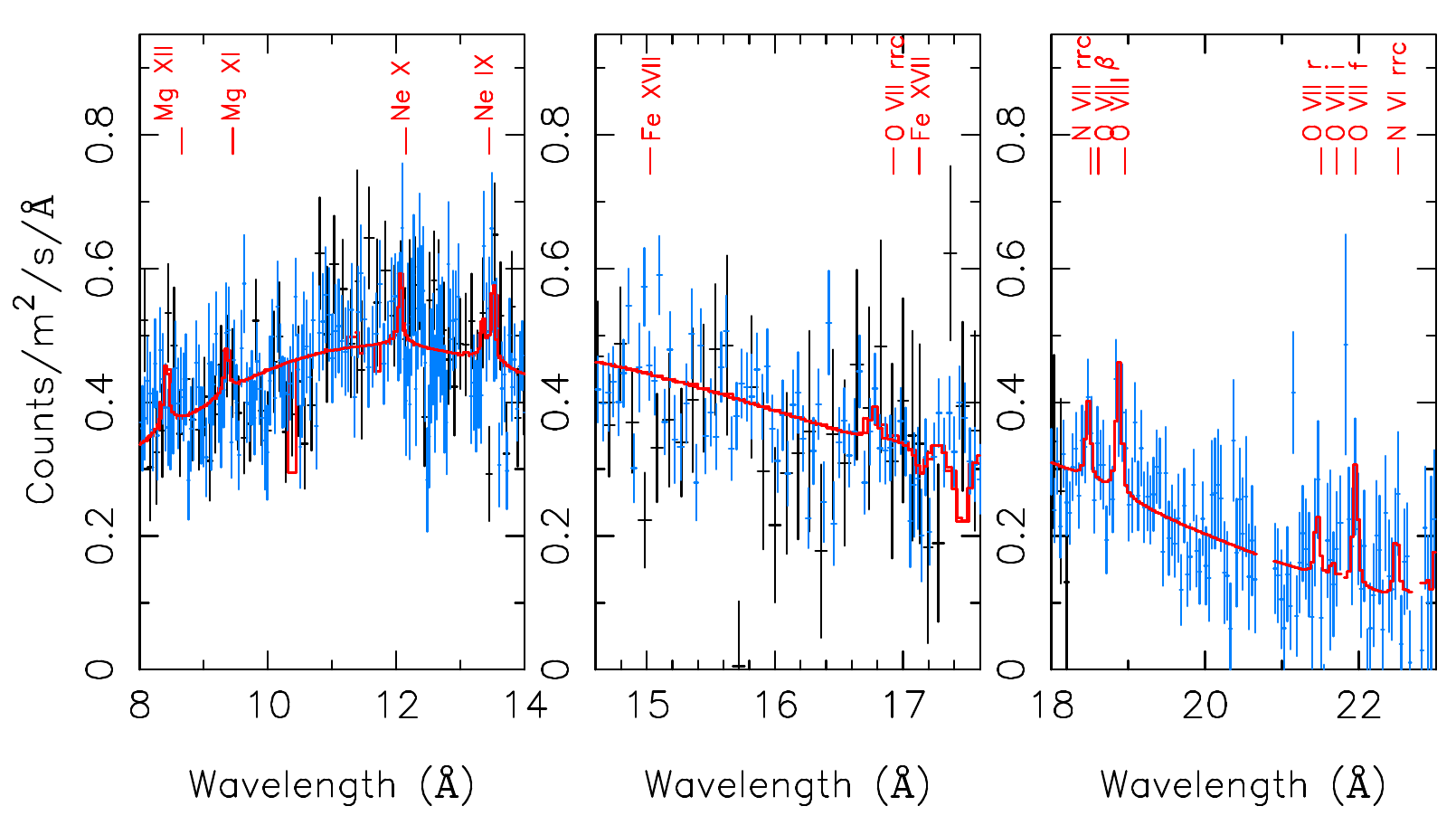}
   \vspace{-0.5cm}
      \caption{XMM-{{Newton}} RGS time-averaged spectrum (order 1 in blue, order 2 in black) and gaussian emission lines: zooming on the Mg-Ne K, Fe L and O K edges. Notice the weakness of Fe lines, the stronger {\ovii} forbidden line and the tentative RRC profiles (hints of photoionisation).}
         \label{fig:spec_all_gaus_lines}
   \vspace{-0.3cm}
   \end{figure}

The single-trial significance was computed as the squared root of the $\Delta C$-stat improvement with respect to the continuum and was then multiplied by the sign of the gaussian normalisation to distinguish between emission- and absorption-like features. The line scan gets rid of some noise in the plot and improves visualisation of possible spectral features including those in emission near the 1s-2p transitions of {\oviiviii} and {\neixx}. These were previously found in the scan of the archival observation id: 0655050101 (P17) and in other ULXs (such as NGC 1313 ULX-1 and NGC 247 ULX-1, \citealt{Pinto2016nature,Pinto2021}) although some clearly show evidence of Doppler shifts (see Fig.\,\ref{fig:spec_all_gaus}). Further absorption-like features were found between 0.7-0.8 keV and above 1 keV just as in the archival observation id: 0655050101 (P17). Further, negative residuals can be seen between 3-4 keV in the EPIC spectra, away from any significant instrumental features in the mirror response, likely from blueshifted Si-S lines.

We attempted a quick modelling of the dominant emission lines with simple gaussian profiles since they appear very narrow and close to their laboratory wavelengths (see Fig.\,\ref{fig:spec_all_gaus_lines}). The centroids of the weak Lyman $\alpha$ (Ly$\alpha$) lines from {\mgxii} and {\nex} and the He-like triplets from {\mgxi} and {\neix} are consistent with their rest values. The {\fexvii} lines, which are often signatures of collisional plasma, are not significantly detected. The lines from cooler plasma, e.g. {\oviii} Ly$\alpha$ and {\ovii} triplet are blueshifted by about {$-1900\pm200$ km/s} as already found in P17, hinting at a slow outflow. Evidence of radiative recombination continua (RRC, another signature of photoionisation) is shown by weak lines in proximity of the wavelengths expected for the abundant, recombining ions ({\ovii} and {\nvivii}). In particular, the presence of a {\nvii} RRC is suggested by the presence of a rather strong line at 18.5-18.6\,\AA\ which is difficult to explain only with {\ovii}\,$\beta$ line emission given that the measured flux is comparable to that of the {\ovii}\,$\alpha$ resonant line at 21.6\,\AA. The profile of these emission lines is not fully resolved due to their width being less than 1000 km/s. Tighter constraints were obtained by fitting multiple lines with a physical model (see Sect.\,\ref{sec:model_scan}). The relative ratios of He-like lines can be used to retrieve useful information on the plasma state, as is shown in Sect.\,\ref{sec:discussion}.

\subsubsection{Physical model scan}\label{sec:model_scan}

In order to identify the nature of the spectral features and correctly estimate any Doppler shift, we performed a modelling through the aid of optically-thin thermal plasma. We tested emission and absorption models of plasma in collisional ionisation (CIE) or photoionisation (PIE) equilibrium but focusing mainly on the latter given the nature of the source and the variability of the X-ray spectral lines seen in other ULXs (see e.g. \citealt{Pinto2020b,Pinto2021}; more detail in Sect.\,\ref{sec:individual_obs}). As we potentially expect large Doppler shifts of the plasma, to find the global best spectral fit, we performed a deep exploration of the parameter space for the two most relevant parameters: the Doppler shift or velocity projected along the line of sight ($v_{\rm LOS}$) and the plasma state parameter (the temperature, k$T$, for CIE and the ionisation parameter $\xi=L_{\rm ion}/n_{\rm H}R^2$ for PIE, where $L_{\rm ion}$ is the ionising luminosity, $n_{\rm H}$ the volume density, and $R$ the distance of the plasma from the ionising source).

{CIE - line emission.} In the case of line-emitting CIE plasma similar to a collimated outflow or jet a-la SS433 \citep{Marshall2002}, we adopted the \texttt{cie} component in {\scriptsize{SPEX}} which calculates the emission spectrum of a plasma in collisional equilibrium. The \texttt{cie} component was multiplied by the redshift component \texttt{reds} in {\scriptsize{SPEX}} to account for Doppler shifts. We then compute ($v_{\rm LOS}$, k$T$) \texttt{cie} model grids and leave the normalisation (or emission measure) free to vary. We adopted a logarithmic
grid of temperatures between 0.1 and 5 keV (50 points), and a linear grid of line-of-sight velocities, $v_{\rm LOS}$, 
between $-0.3c$ (blueshifted jet) and $+0.3c$ (redshifted jet). \textcolor{black}{{We tested a velocity dispersion $v_{\sigma}$ of 100, 1000, 5000 and 10000 km/s with LOS velocity steps of 300, 700, 1000 and 1500 km/s, respectively, to speed up the computation as done for the gaussian line scan, see  Sect.\,\ref{sec:gaussian_scan} and \citealt{Kosec2018b}}).} {The k$T$ range was chosen below 5\,keV because any hotter component would \textcolor{black}{produce the dominant lines} above 5 keV, i.e. in the region where EPIC BKG starts to matter and out of the RGS reach.} \textcolor{black}{Moreover, the RGS energy range is too short to distinguish weak lines emitted by plasma with k$T$ of a few keV or larger (see, e.g., \citealt{Lepore2025}).}

Given the limited number of counts for individual lines and the integration time and the variability of the line centroid, 
which could smear out the triplets when stacking all the spectra, we do not expect to obtain strong constraints on the relative ratios of the He-like triplets e.g. those of {\ovii} and {\neix}.
This implies loose constraints on the plasma volume (or number) density, whose fitting may result in an unnecessary increase in the computation time.
We therefore chose not to fit the volume density and adopted $n_{\rm H} = 10^{10}$ cm$^{-3}$,
which is a lower limit found for NGC 1313 ULX-1 \citep{Pinto2020b} and agrees with the measured {\ovii} line ratios (see Sect.\,\ref{sec:emitting_plasma}).
This might only slightly affect the overall flux and column density of the line-emission component.

\begin{figure}
   \centering
   \includegraphics[scale=0.45]{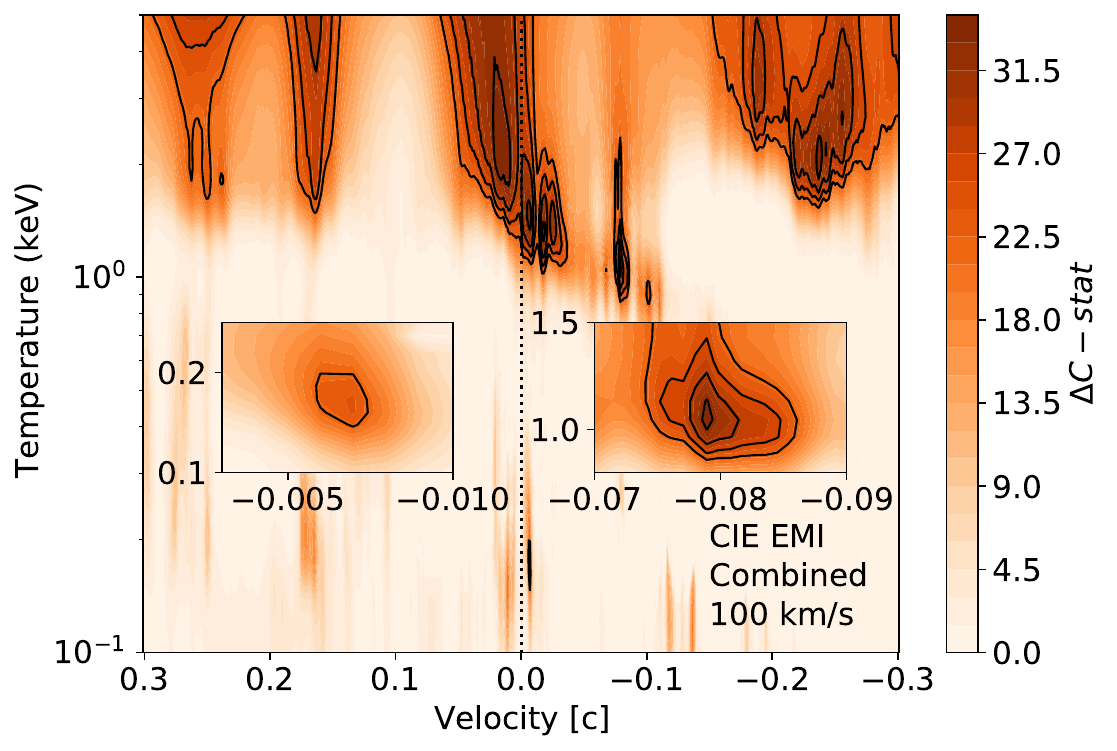}
   \includegraphics[scale=0.45]{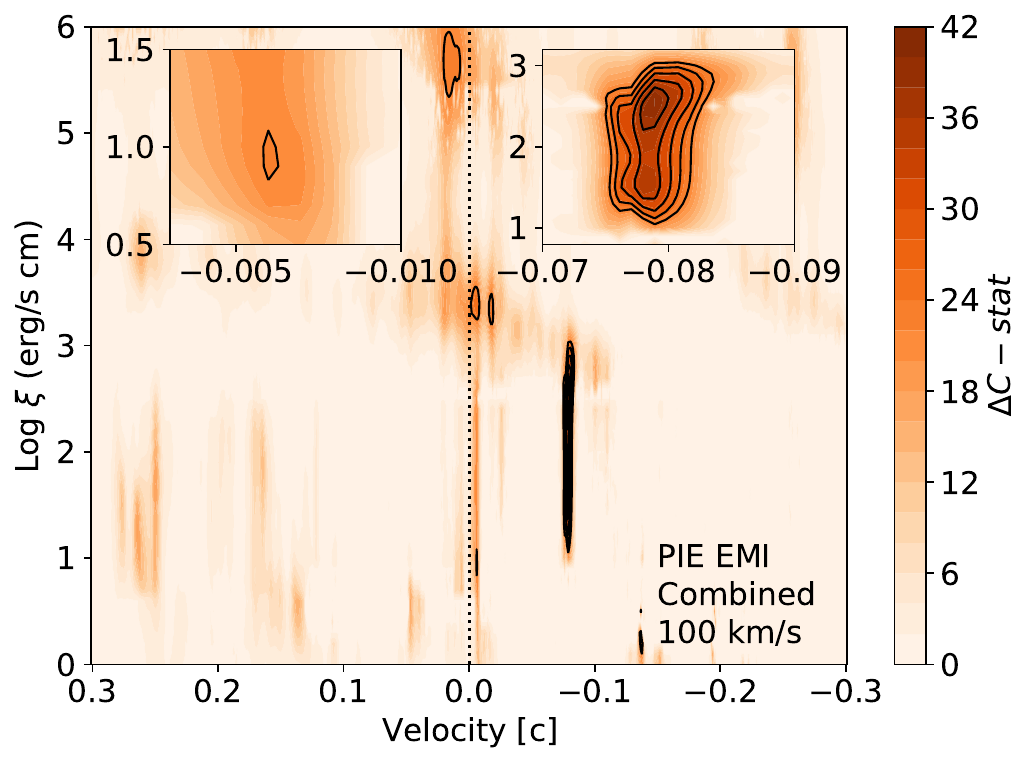}
   \includegraphics[scale=0.45]{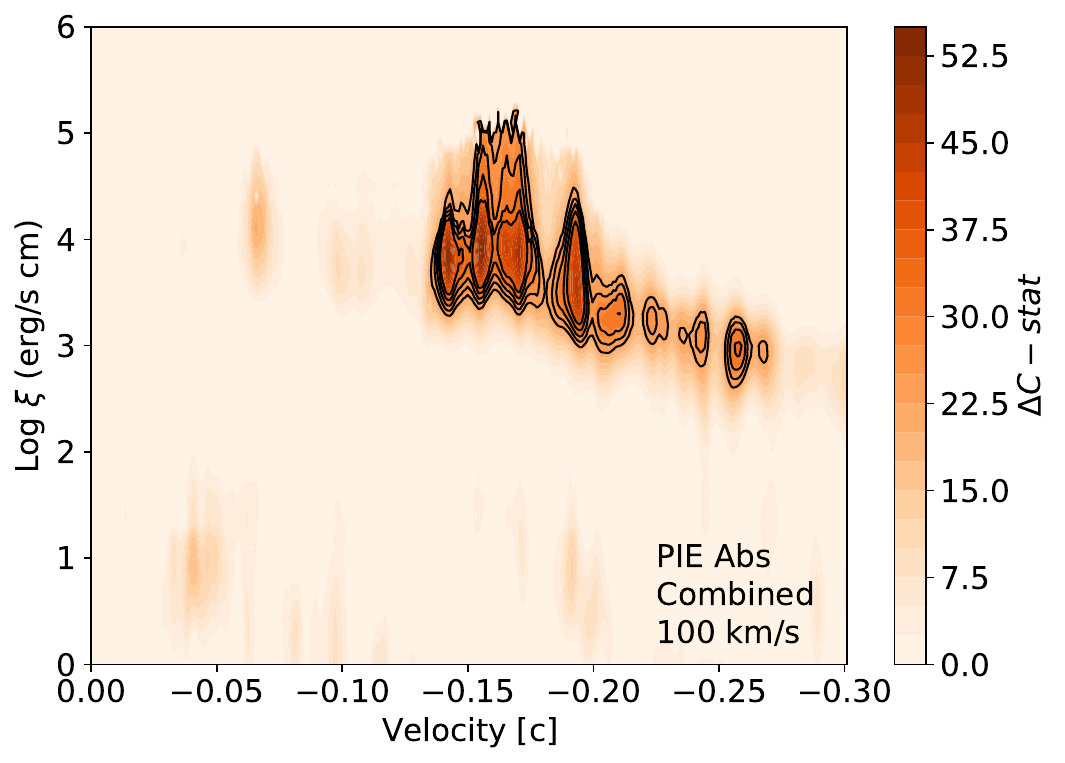}
   \vspace{-0.1cm}
      \caption{XMM-Newton time-averaged grids of physical models. From top to bottom: CIE (\texttt{cie} model) emission, PIE (\texttt{pion}) emission and  (\texttt{xabs}) absorption. The black contours refer to significance levels from 3 to $5\,\sigma$ \textcolor{black}{with steps of $0.5\,\sigma$ estimated with MC} simulations (Sect.\,\ref{sec:monte_carlo_simulations}).}
         \label{fig:spec_tavg_grids}
   \vspace{-0.1cm}
   \end{figure}

A loop is computed throughout these model grids and is shown as a contour plot for $v_{\sigma}$ = 100 km/s in Fig.\,\ref{fig:spec_tavg_grids} (top panel). Multiple solutions appear with temperatures around 1-2 keV and different $v_{\rm LOS}$ (a few thousands km/s very similar to NGC 247 ULX-1, $-0.08c$ and $-0.2c$ previously reported in P17 for obsid 0655050101). The overall $\Delta C$-stat is well above 30, which according to previous work already indicates a highly significant detection, although strong degeneracy seems to be present for this particular model. 

We did not test CIE absorption plasma as it is not expected to dominate the absorption measure distribution, particularly in the presence of a strong X-ray radiation source (the ULX itself). Moreover, with the present spectral quality and signal-to-noise ratio, it would be challenging to distinguish CIE from PIE absorption models because they both mainly produce absorption lines related to the electron transitions from the resonant levels.

{PIE - line emission.} As mentioned before, the input SED for the photoionisation balance computation was extrapolated from the best-fit continuum model ({see Appendix\,\ref{appendix:thermal_stability} and} Fig.\,\ref{fig:sed_ionbal}). The \texttt{pion} model in {\scriptsize{SPEX}}  can be used to produce spectra of PIE plasma both in emission and absorption depending on the adopted values for the solid angle $\Omega/4\pi$ and the covering fraction $f_{\rm cov}$ which vary between 0 and 1 by definition. In the case of emission-line spectra, we fixed $\Omega/4\pi=1$ and $f_{\rm cov}=0$ to speed up the computation. We scanned the time-averaged EPIC+RGS spectra with the grids of \texttt{pion} models (on top of the continuum model) with the same multi-dimensional routine used for the \texttt{cie} model, and a similar parameter space (actually identical for the Doppler shift).
We adopted a logarithmic grid of ionisation parameters (log\,$\xi$ [erg/s cm] between 0 and 6 
with 0.1 steps).
The only free parameter for \texttt{pion} is the column density, $N_{\rm H}$. As previously done for the \texttt{cie} grids, we adopted $n_{\rm H} = 10^{10}$ cm$^{-3}$ for the \texttt{pion} model grids. Unlike the \texttt{cie} grids, the \texttt{pion} grids show a clear absolute maximum with velocity $-0.08c$, i.e. an outflow, and a significantly higher $\Delta C$-stat of 42 (see Fig.\,\ref{fig:spec_tavg_grids}, middle panel). Further local minima were found at outflow velocities slightly above $0.1c$ and close to the restframe, likely pointing towards a multiphase or variable plasma as suggested in P17 or possible degeneracy between different solutions.

The CIE and PIE solutions at low temperature (k$T_{\rm e}\lesssim0.2$ keV or log\,$\xi\sim 1$) and close to the restframe ($|v_{\rm LOS}|\lesssim0.01c$) are primarily responsible for fitting the {\oviiviii} (and weaker {\neixx}) features close to rest, while those with intermediate temperature and velocity ($-0.08c$) are trying to describe the jump around 1 keV (as previously suggested in P17, albeit with greater significance in the combined data).

{PIE - line absorption.} For a fast execution of the parameter space exploration through grids of photoionised absorbing plasma (which is later crucial for expensive Monte Carlo simulations), we used the fast \texttt{xabs} model in {\scriptsize{SPEX}}. This adopts the ionisation balance pre-calculated through the {\scriptsize{SPEX}} task \texttt{xabsinput} once the SED is provided in input. The ionisation balance files used by \texttt{xabs} include the conversion of the ionisation parameter to the temperature and the ionic fraction for every element at a given value of log $\xi$. The grid of photoionised \texttt{xabs} models were calculated and fitted in the same way as the \texttt{pion} models, but assuming line-of-sight velocities, $v_{\rm LOS}$, ranging between -0.3c and zero (i.e. only restframe or outflowing plasmas in absorption because we do not expect extreme inflows). The \texttt{xabs} results are shown in Fig.\,\ref{fig:spec_tavg_grids} (bottom panel). Aside from some degeneracy between multiple solutions (likely driven by spectral stacking), it is obvious that the absorbing plasma flows out with a relativistic speed of $-0.15c$ or higher, as found for the archival observation in P17. The very large $\Delta C$-stat, well above 50, suggests a very high confidence level, most likely above $5\,\sigma$ (see below).

Given the particularly large improvement in the spectral fits and the presence of degeneracy due to the stacking process, we performed the final fits (with the addition of both emission and absorption components) only on the spectra of the individual observations (see below). \textcolor{black}{Moreover, the multiple solutions are a characteristic of the combined spectrum. In the individual spectra only one solution is detected (if detection actually occurs, see Sect.\,\ref{sec:individual_obs}). Given that a meaningful non-degenerate modelling aimed at comparing variability over time can be performed only on individual observations, we refrained from the test of more complicated emission models for the combined spectrum.}

\subsection{Individual observations}\label{sec:individual_obs}

For individual observations, we repeated the several steps performed for the time-averaged spectrum. One main difference is that we do not include the $2^{\rm nd}$ order spectra of individual observations; their low statistics would result in an unnecessary increase in computing time. Our steps were the following:
\begin{itemize}
    \item Estimation of the continuum with the double thermal model as done in Sect.\,\ref{sec:continuum_and_SED}. The spectra of individual observations are shown in Fig.\,\ref{fig:spec_continuum}. 
    \vspace{0.1cm}
    \item Performance of a gaussian line scan for the individual spectra following the same approach as in Sect.\,\ref{sec:gaussian_scan}. The results of the scan are shown in the lower panels of Fig.\,\ref{fig:spec_all_gaus}.
    \vspace{0.1cm}
    \item Multi-dimensional scanning of the spectra with different physical models (emission by collisional or photoionised plasmas and absorption by a photoionised plasma) with the same routine used in Sect.\,\ref{sec:model_scan}. The results of the CIE and PIE scans are shown in Fig.\,\ref{fig:individual_scans}. The black contours refer to the confidence level estimated through Monte Carlo simulations as described in the following section.
    {Diagnostic checks for secondary peaks, alternative solutions and degeneracy are performed adopting the spectra from obsid 0655050101 (see Appendix\,\ref{appendix:individual_fits} and Fig.\,\ref{fig:secondary_scans}).}
    \vspace{0.1cm}
    \item
    For the individual observations, we also produced plots for the best-fit continuum+wind model. Here, on top of the double thermal continuum, we also added a combination of emitting (\texttt{pion}) and absorbing (\texttt{xabs}) photoionised plasma components (see Fig.\,\ref{fig:spec_bestfit} and Table\,\ref{table:wind_properties}). 
    {In Appendix\,\ref{appendix:spectral_components} and Fig.\,\ref{fig:spectral_components}, we show the zoom on the RGS spectra highlighting the contributions of the main model components.}
    \item 
    \textcolor{black}{As a diagnostic check, we substituted the \texttt{xabs}  with the more accurate \texttt{pion} model in absorption (assuming $\Omega/4\pi=0$ and $f_{\rm cov}=1$) and refit the spectra starting from the best fit obtained with \texttt{xabs} (see Table\,\ref{table:wind_properties}). No significant change was observed and the uncertainties did not decrease with \texttt{pion}, for which we preferred to keep the results obtained with \texttt{xabs} to be also consistent with the simulations performed.}
\end{itemize}
 
\subsection{Monte Carlo simulations and detection significance} \label{sec:monte_carlo_simulations}

The significance of the line detection was obtained by running identical searches in the same parameter space onto a large number of simulated featureless spectra. Following the code developed in \citet{Pinto2020b}, we performed Monte Carlo (MC) simulations of a physical model scan (PIE / \texttt{xabs}) onto 16,000 spectra that were simulated adopting the best-fit continuum model of the RGS+EPIC obsid 0655050101. The results are shown in Fig.\,\ref{fig:mc_simulations}. A histogram of $\Delta C$-stat occurrences was produced to estimate the significance of the detection in real data. Given the stability of the histogram slope ($-0.26$) above about 5,000 simulated spectra (as already pointed out in \citealt{Pinto2021}) a forecast was made for $1.5\times10^{5}$ and $1.7\times10^{6}$ simulations to enhance the significance sensitivity up to 4.5 and 5.0\,$\sigma$. The other observations have comparable exposure times, flux, and number of counts implying that the results for obsid 0655050101 can be extended to the other ones. In summary, the detections in the data that refer to $\Delta C$-stat improvements above 22, 29 and 36 with respect to the continuum-only model correspond to confidence levels above 3, 4 and 5\,$\sigma$, respectively. The confidence levels can be found as black contours in Fig.\,\ref{fig:individual_scans}. 
This exercise was repeated for the time-averaged spectrum (to quantify the significance of the detections in Fig.\,\ref{fig:spec_tavg_grids}) for five thousand simulations. The histogram of the $\Delta C$-stat occurrences has a comparable slope (about $-0.25$) and agrees with that from the MC simulations performed for NGC 247 ULX-1 and similar cases (e.g. \citealt{Pinto2021, Kosec2021, Gu2022}).

Final spectral fits were performed for the time-averaged and individual spectra. We tested two combinations of optically-thin plasmas: a hybrid one with emitting CIE and absorbing PIE (\texttt{cie} + \texttt{xabs}) and a purely photoionised emitting plus absorbing (\texttt{pion} + \texttt{xabs}) plasma. The properties of the \texttt{xabs} did not change within the uncertainties given the narrow width of the lines. For the same reason, $\Delta C$ did not decrease significantly while fitting multiple components (they are not mutually dependent; see also the discussion in \citealt{Pinto2021}). In Table\,\ref{table:wind_properties} we report the best-fit values, the spectral improvement, and the estimated significance of each component derived using the MC simulations. We also quote the bolometric luminosities once they were corrected for absorption from the \texttt{xabs} component. In addition, we also report the spectral hardness computed as the ratio between the two luminosities computed in the 2--10 keV and 0.3--2 keV energy bands {(estimated with the continuum model fits)}. Finally, the root-mean-square estimated as the square root of the excess variance in the 0.3-10 keV light curves shown in B22 is also reported in the same table.

\begin{table*}
\begin{center}
   \vspace{-0.1cm}
\caption{Best-fit plasma parameters in each EPIC+RGS spectral fit.}  
\label{table:wind_properties}     
\renewcommand{\arraystretch}{1.2}
 \small\addtolength{\tabcolsep}{4pt}
 \vspace{-0.1cm}
\scalebox{0.95}{
\hskip-0.0cm\begin{tabular}{@{}ccccccc}     
\hline  
Parameter	 & Combined		& 0655050101		& 0824570101		& 0864810101		& 0883960101 \\
\hline
$L_{\rm BOL}$ ($10^{39}$ erg/s) & 1.68 $\pm$ 0.03		& 1.71 $\pm$ 0.03 	& 1.26 $\pm$ 0.05 	& 1.69 $\pm$ 0.05	& 2.08 $\pm$ 0.05 \\
$L_{\rm BOL}^{\rm corr}$ ($10^{39}$ erg/s) & 2.58 	& 2.92	& 1.70 	& 2.75 	& 2.33 \\
Hardness &  0.097 $\pm$ 0.002 	& 0.092 $\pm$ 0.004	& 0.102 $\pm$ 0.004	& 0.103 $\pm$ 0.004	& 0.089 $\pm$ 0.004 \\
RMS {(\%)} &  --- 	& 12.3 $\pm$ 0.3	& 3.73 $\pm$ 0.01	& 15.34 $\pm$ 0.04	& 7.79 $\pm$ 0.01 \\
\hline
$v_{\rm LOS}^{\rm xabs}$ $(c)$ & -0.156 $\pm$ 0.001	& -0.176 $\pm$ 0.001	& -0.141 $\pm$ 0.002	& -0.107 $\pm$ 0.001	& -0.046 $\pm$ 0.001 \\
$v_{\sigma}^{\rm xabs}$ (km/s) & 100 $\pm$ 30		&  50 $\pm$ 50		& 250 $\pm$ 250	& 100 $\pm$ 50		& 480 $\pm$ 480 \\
$\log \xi^{\rm xabs}$ (erg/s cm) & 4.06 $\pm$ 0.09		& 3.94 $\pm$ 0.15	& 3.81 $\pm$ 0.15	& 3.97 $\pm$ 0.17	& 1.26 $\pm$ 0.17 \\
$N_{\rm H}^{\rm xabs}$ ($10^{24}$/cm$^2$) & 0.65 $\pm$ 0.21		& 0.62 $\pm$ 0.25	& 0.33 $\pm$ 0.12	& 0.55 $\pm$ 0.31	& 0.06 $\pm$ 0.03 \\
$\Delta C$	& 53 & 27 & 26 & 21 & 22 \\
$CL \, (\sigma)$ & $>5$ & $3.6$ & $3.5$ & $2.9$ & $3.0$ \\
\hline
$v_{\rm LOS}^{\rm pion}$ $(c)$ & -0.082 $\pm$ 0.001	& -0.006 $\pm$ 0.001	& -0.079 $\pm$ 0.001 & 0.009 $\pm$ 0.001	& -0.140 $\pm$ 0.001 \\
$v_{\sigma}^{\rm pion}$ (km/s) & 400 $\pm$ 400		& 100 $\pm$ 100	& 100 $\pm$ 100	& 700 $\pm$ 560	& 500 $\pm$ 500 \\
$\log \xi^{\rm pion}$ (erg/s cm) & 2.5 $\pm$ 0.1		& 0.94 $\pm$ 0.1	& 1.1 $\pm$ 0.1 		& 0.25 $\pm$ 0.05	& 0.05 $\pm$ 0.05 \\
$L_{\rm X}^{\rm pion}$ ($10^{38}$ erg/s) & 0.60 $\pm$ 0.09		& 0.83 $\pm$ 0.17	& 0.42 $\pm$ 0.12	& 1.21 $\pm$ 0.28	& 0.81 $\pm$ 0.17 \\
$\Delta C$	& 42 & 36 & 17 & 24 & 27 \\
$CL \, (\sigma)$ & $>5$ & $4.8$ & $<2.5$ & $3.3$ & $3.6$ \\
\hline
$v_{\rm LOS}^{\rm cie}$ $(c)$ & -0.004 $\pm$ 0.017	& -0.006 $\pm$ 0.001	& -0.030 $\pm$ 0.004 & 0.028 $\pm$ 0.006	& -0.136 $\pm$ 0.001 \\
$v_{\sigma}^{\rm cie}$ (km/s) & $7000 \pm 3000$	& 350 $\pm$ 350	& 1000 $\pm$ 1000	& 4074 $\pm$ 2600	& 342 $\pm$ 342 \\
k$T_{\rm e}^{\rm cie}$ (keV) & 1.4 $\pm$ 0.1		& 0.18 $\pm$ 0.02	& 2.05 $\pm$ 0.23	& 0.11 $\pm$ 0.02	& 0.07 $\pm$ 0.01 \\
$L_{\rm X}^{\rm cie}$ ($10^{38}$ erg/s) & 1.53 $\pm$ 0.32		& 0.90 $\pm$ 0.28	& 0.92 $\pm$ 0.20	& 2.53 $\pm$ 1.12	& 1.86 $\pm$ 0.99 \\
$\Delta C$	& 41 & 26 & 22 & 21 & 22 \\
$CL \, (\sigma)$ & $>5$ & $3.5$ & $3.0$ & $2.9$ & $3.0$ \\
\hline             
\end{tabular}}

\vspace{0.3cm}

   \vspace{-0.1cm}
Notes: Bolometric luminosities are computed within $10^{-4} - 10^{\,2}$ keV. The root-mean-square (RMS, in \% units) has been computed using the 0.3--10\,keV EPIC-pn light curves (time bin size of 1\,ks, taken from B22). $L_{\rm BOL}^{\rm corr}$ indicates the luminosity corrected for photoionised \texttt{xabs} absorption. The spectral hardness {(estimated from continuum-only fits)} yields the ratio between the hard and soft X-ray luminosities ($L_{\rm 2-10 \, keV} / L_{\rm 0.3-2 \, keV}$). The X-ray luminosities of the plasma emitting components refer to the 0.3-10\,keV band. The $\Delta C$ show the fit improvement; the confidence level ($CL$) expressed in $\sigma$ units is computed through Monte Carlo simulations (see Sect. \ref{sec:monte_carlo_simulations}) and has an uncertainty of about $0.1\sigma$.
\vspace{-0.3cm}
\end{center}
   \vspace{-0.1cm}
\end{table*}

\section{Discussion}\label{sec:discussion}

{In this section we address the nature of the optically-thin plasma components and place some constraints on the nature of the accretor and its energetic feedback on the surrounding medium.}

\subsection{The nature of the line emitting plasma} \label{sec:emitting_plasma}

{The model grids have shown multiple solutions when we adopted CIE plasma and low velocity dispersion (see Fig.\,\ref{fig:spec_tavg_grids}), while a better fit was obtained with a very large broadening ($v_{\sigma}\sim7000$ km/s, see Table\,\ref{table:wind_properties}). This means that the \texttt{cie} struggles to describe all features and that some solutions are degenerate.}
    
\begin{figure}
   \centering
   \includegraphics[scale=0.4]{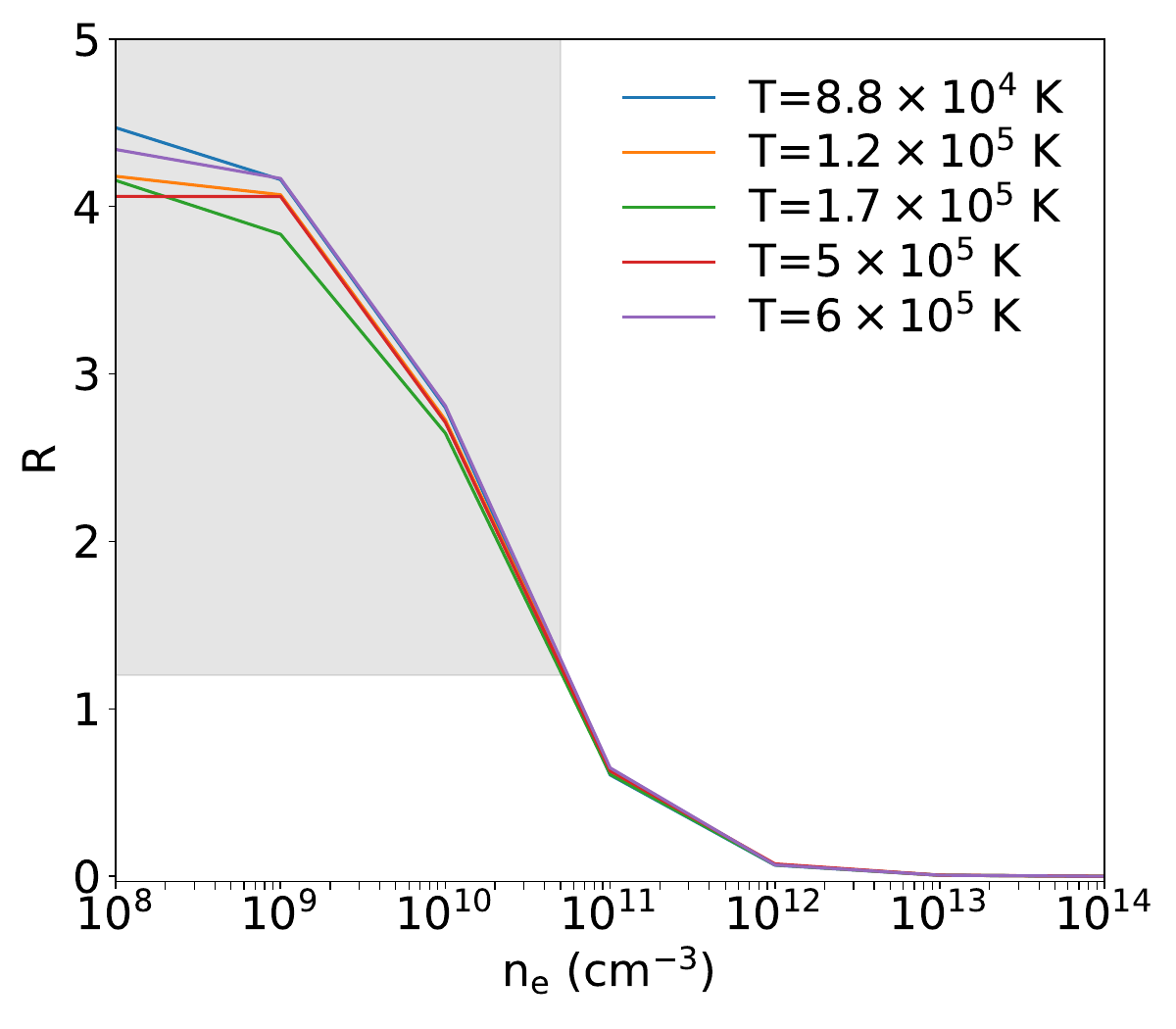}
   \includegraphics[scale=0.4]{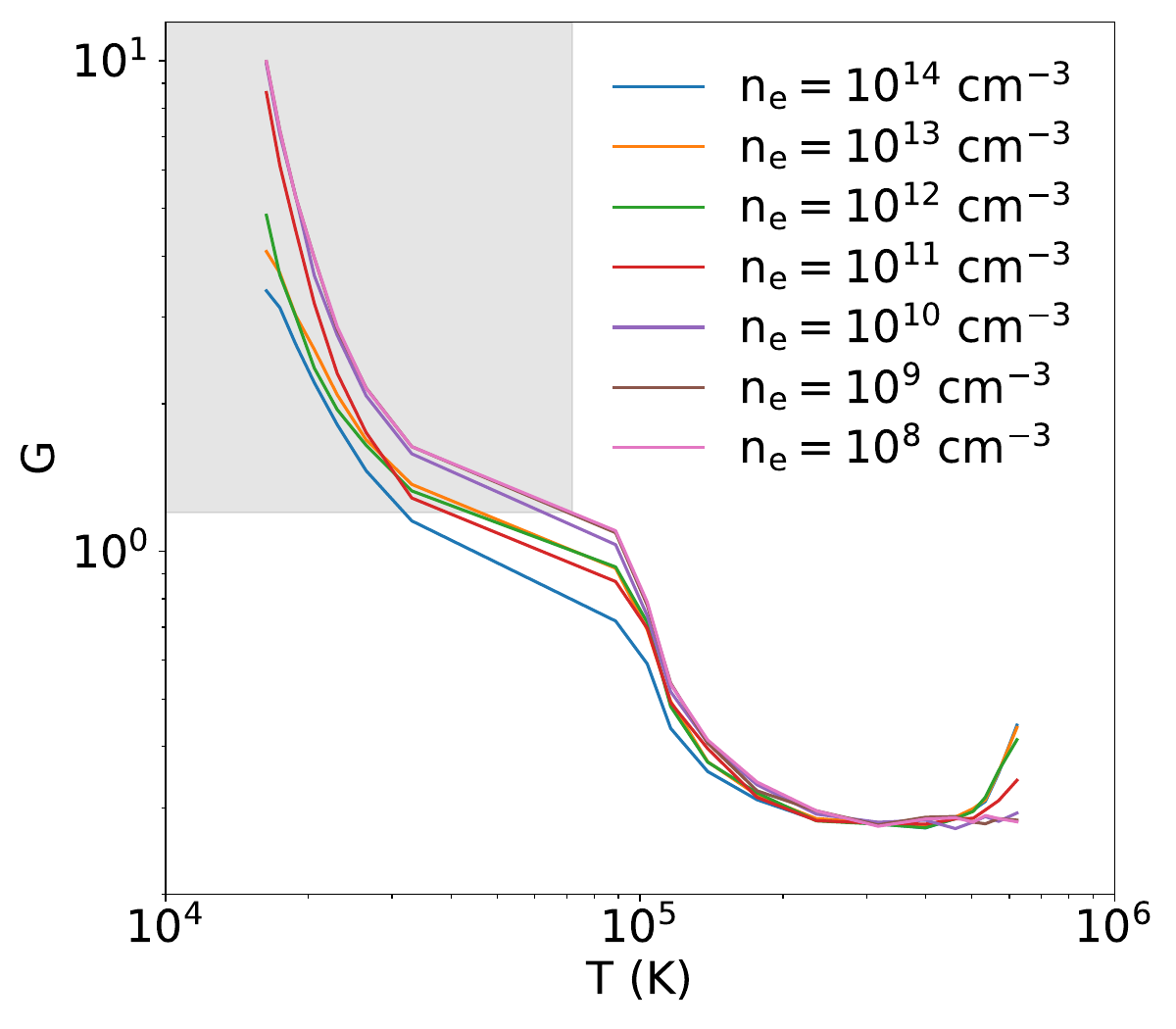}
   \vspace{-0.1cm}
      \caption{{\ovii} He-like triplet $R$ (top) and $G$ (bottom) ratios curves calculated from the time-averaged SED. {Grey shaded areas indicate the regions compatible with the results from the combined spectrum and the upper limits on the temperature and the density} (see Sect.\,\ref{sec:discussion}).}
         \label{fig:g_r_ratios}
   \vspace{-0.1cm}
   \end{figure}

    The $G-R$ line ratios, measured for the relevant He-like triplets such as those of {\ovii} and {\neix}, may be useful to place constraints on the plasma density and temperature (see, e.g., \citealt{Porquet2000}). The time-average RGS spectrum provides by far the best statistics and allows gaussian modelling of the individual lines (see Fig.\,\ref{fig:spec_tavg} and \ref{fig:spec_all_gaus_lines}). Using the partly-resolved {\ovii} triplet (resonant $r$, intercombination $i$ and forbidden $f$ lines at 21.6, 21.8, and 22.1{\,\AA}, respectively), {we obtain two lower limits $R=f/i>1.2$ (at $68\,\%$, with most probable value around 4) and $G=(f+i)/r>1.2$ (at $68\,\%$, with most probable value 4.6).}
    We used {\scriptsize{SPEX}} \texttt{pion} code to compute the $R$ and $G$ ratios for the {\ovii} triplet as functions of the electronic density, $n_{\rm e}$, and the electronic temperature $T$, respectively. As broadband SED we adopted the best-fit continuum model for the time-averaged spectrum. In output, through the \textit{ascdump} command, \texttt{pion} provides the relationship between the ionisation parameter and the temperature, and many more info including the fluxes of the lines. We performed the computation through a grid of $\log{\xi}$ between 0 and 4.0 with 0.2 steps, and $n_{\rm e}$ from $10^{8}$ to $10^{14}\,\text{cm}^{-3}$ with a logarithmic step of 10. We also adopted a column density $N_{\rm H} = 10^{22}$ cm$^{-2}$ and $v_{\sigma}=250$ km/s, which agree with the best-fit values. The curves are plotted in Fig.\,\ref{fig:g_r_ratios}. The comparison of such curves with the data measurements implies, through the $G$ ratio, a temperature $kT\lesssim10^5$\,K (consistent with the \texttt{pion} best-fit results, see also Fig.\,\ref{fig:sed_ionbal}).
    The $R$ ratio instead yields a number density around $7\times10^{10}\text{cm}^{-3}$ (or an upper limit of $10^{11}\text{cm}^{-3}$). This suggests that we are looking at a photoionised plasma with a slightly lower density than that seen in Eddington-limited Galactic XRB winds or accretion disc coronae {(see, e.g., \citealt{Psaradaki2018,Neilsen2023} and references therein)}. The RGS resolution at 13\,{\AA} and signal-to-noise ratio are too low to place significant constraints on the {\neix} triplet.

    {From the definition of the ionisation parameter we can resolve for the density in order to obtain a constraint on the radius, i.e., the distance of the photoionised line-emitting plasma from the inner portion of the accretion disc, $R=\sqrt{L_{\rm ion} / (n_{\rm H} \xi)}$. Assuming the upper limit of the density mentioned above and the {\texttt{pion}} best-fit parameters determined in the time-averaged spectrum (see Table\,\ref{table:wind_properties}), we obtain a lower limit on the distance {$R\gtrsim7.3\times10^{12}$\,cm $\sim 0.5$\,AU (or $5\times10^6\,R_{\rm G}$ for a $10\,M_{\odot}$ black hole).} This is similar to what was obtained for NGC 1313 ULX-1 \citep{Pinto2020b}. The semi-major axis of the orbits for similar systems such as SS 433 and NGC 7793 P13 is of the order of $0.4-0.8$\,AU \citep[][the Roche lobe would be even smaller]{Bowler2011,Fuerst2018}, which would indicate a location further out than the outermost part of the accretion disc, that is expected from an equatorial, thermal, wind. The entire binary might be engulfed in an optically-thin line-emission envelope with a large solid angle, just as could occur in SS 433 \citep{Waisberg2019}. This slower outflow component might be associated with the broad H$\alpha$, consistent with emission from a hot disc wind (600 km/s, \citealt{Zhou2023}).}

    \subsection{A multiphase outflow}
    The presence of plasmas moving at about {$\pm1000$ km/s and $-0.1c$ (in emission) and around $-0.15c$} (in absorption) points towards a stratified, likely multiphase, outflow. {We also note that the \texttt{pion} modelling of the combined spectrum favours a fast {$-0.08c$} outflow with an ionisation parameter higher ($\log\xi=2.5$) than the individual fits due to a secondary solution that is present in all scans and sticks out once all data are taken into account (see Table\,\ref{table:wind_properties}).} The fastest components provide evidence for mildly-relativistic matter driven by the strong radiation pressure. In fact, their velocities and temperatures agree with the predictions from theoretical simulations of super-Eddington accretion (see, e.g., \citealt{Ohsuga2005,Takeuchi2013}). Such high Doppler velocities with a corresponding low velocity broadening are also in line with the recent results obtained with the high-resolution Fe K spectra of highly-accreting supermassive black holes taken with the \textit{XRISM / Resolve} microcalorimeter (see, e.g., \citealt{Xu_2025_PDS456, XRISM_2025_PDS456}). The low-velocity, low-temperature plasma responsible for the narrow {\oviiviii} emission lines is most likely a result of thermal pressure and ionisation of the outer regions of the disc (a.k.a. thermal winds; see, e.g., \citealt{Pinto2023a}). Using the ionisation balance (k$T_{\rm e}-\log\xi$ curves) computed with \texttt{pion}, we estimated a temperature above $10^4$\,K and a ratio of about 10 between the radiation pressure ($F/c$) and the thermal pressure ($n_{\rm H}kT$), which is given by $\Xi = F / n_{\rm H} c kT = 19222 \, \xi / T$, with $F = L / 4 \pi R^2$ \citep{Krolik1981} (see Fig.\,\ref{fig:sed_ionbal}). {Such values agree} with the expectation of the theory of thermal winds (see, e.g., \citealt{Middleton2022}).

    A comparison is performed between the plasma properties to address its thermal stability, in particular, through the $S$ curve. The $S$ curve is described as the relationship between the state parameter (k$T$ or $\xi$) and the radiation / thermal pressure ratio $\Xi$ (see Fig.\,\ref{fig:sed_ionbal}). Neither the \texttt{pion} nor the \texttt{xabs} best-fit $\xi$ values are found in correspondence of negative slopes or unstable branches {with the exception of the absorbing plasma ({$v_{\rm LOS}=-0.05c$} and $\log\,\xi=1.26$) exhibited in the highest-flux observation (0883960101)} {and, at a lower-significance, in the first, intermediate-flux observation. This} would indicate a sporadic event or even a `failed wind' (e.g., \citealt{POP2021,Miller2025}). The presence of an instability branch between $\log\xi=1.1-1.2$ confirms that the slow/cool and fast/hot plasma components are not in thermal balance with each other and that the outflow is multiphase. {In other words, these two (line-emitting and absorption-line) components do not share the same origin, e.g., with the slow one produced by the outer disc photosphere and the fast one from the inner, super-Eddington, accretion disc. {Alternatively, the former might have been from the inner disk and then slowed and cooled down to the lower branch due to other physical phenomena.}} 

    {The last observation also shows a tentative secondary solution with a $\sim0.25c$ redshift. It corresponds to the \texttt{pion} attempting a fit of a feature at 0.62\,keV (which could also be explained with {\ovii} blueshifted by $-0.08c$) and leftovers residuals in the complex oxygen K edge at 0.5\,keV, which is mainly affected by the cold ISM (and the background). The difficulty to explain such a strong redshift along with the location of the features make us believe that it is most likely a spurious detection. We also note that the primary solution at $-0.140c$ is pegged at its lower bound for both the \texttt{pion} and the \texttt{cie} due to the weakness of lower-ionisation lines and sensitivity at lower temperatures.}

    \subsection{Comparison with other ULXs: an ubiquitous 1 keV line}
    As mentioned in Sect.\,\ref{sec:intro}, high-counts CCD spectra regularly show unresolved spectral residuals around 1\,keV. Among the several proposed scenarios, super-Eddington relativistic winds was a tantalising solution (see, e.g., \citealt{Middleton2014,Middleton2015b}). 
    In \citealt{Pinto2016nature}, the features were resolved for the first time in two ULXs. Simultaneous EPIC+RGS fits indicated that the RGS-detected winds were responsible for most of the EPIC residuals. In soft or supersoft ULX spectra the feature around 1 keV looks like an absorption edge; this is also observed in extragalactic XRBs with $L_{\rm X}\sim10^{39}$ erg/s (see, e.g., \citealt{Urquhart2016,Earnshaw2017}). This would point to evidence of absorption by winds. The results we obtained for SUL NGC 55 ULX-1 and those recently shown for SUL-SSUL NGC 247 ULX-1 show that the feature is actually a combination of high-velocity {($|v|\gtrsim0.15c$) plasma in absorption and low-velocity ($|v|<0.1c$)} plasma in emission (primarily Ne K and Fe L ions with contribution from the {\oviii} edge whose laboratory energy is 0.87 keV, which is shifted above 1\,keV for {$|v_{\rm LOS}|\gtrsim0.15c$)}.
    The line-emitting plasma exhibits a 0.3-10\,keV luminosity ranging between $(0.4-1.2)\times10^{38}$ erg/s, which is of the same order of magnitude as that measured in the RGS spectra of other ULXs although clearly around a factor about 2 lower than in NGC 1313 ULX-1, NGC 5408 ULX-1 and NGC 247 ULX-1 (\citealt{Pinto2016nature,Pinto2020b,Pinto2021}). This could be the result of a higher Eddington ratio and, therefore, stronger radiation pressure in the brighter sources, as noted in a small sample of highly-accreting supermassive black holes \citep{Xu2023}.

    \subsection{The nature of the accretor}
    The highest significance of ultrafast winds in the time-averaged spectrum implies {a long-term sustained radiation pressure and, therefore,} an intrinsic luminosity that is regularly above the critical value, i.e., $L_{cr} \sim 9/4 L_{\rm Edd}$. Moreover, once absorption by the photoionised plasma (\texttt{xabs}) is taken into account, the corrected bolometric luminosity appears to be above $2.5 \times 10^{39}$ erg/s (i.e. 50\% higher). In B22,
    it was suggested that the deviations in the $L-T$ diagram could be interpreted as an increasing contribution of the winds to the soft ($<2$\,keV) X-ray emission at $L_{\rm BOL} \gtrsim L_{cr}$ which would imply a small compact object mass ($5\,M_{\odot}$). If we account for the correction due to photoionised absorption we predict a black hole with a very common mass ($7.5\,M_{\odot}$). If we instead assume that the deviations occur at $L_{\rm BOL} \sim L_{\rm Edd}$, a heavier ($17\, M_{\odot}$) black hole would result, which agrees with arguments related to mass inflow and outflow rates \citep{Fiacconi2017}.

    \subsection{The wind temporal evolution}
\begin{figure}
   \centering
   \includegraphics[scale=0.7]{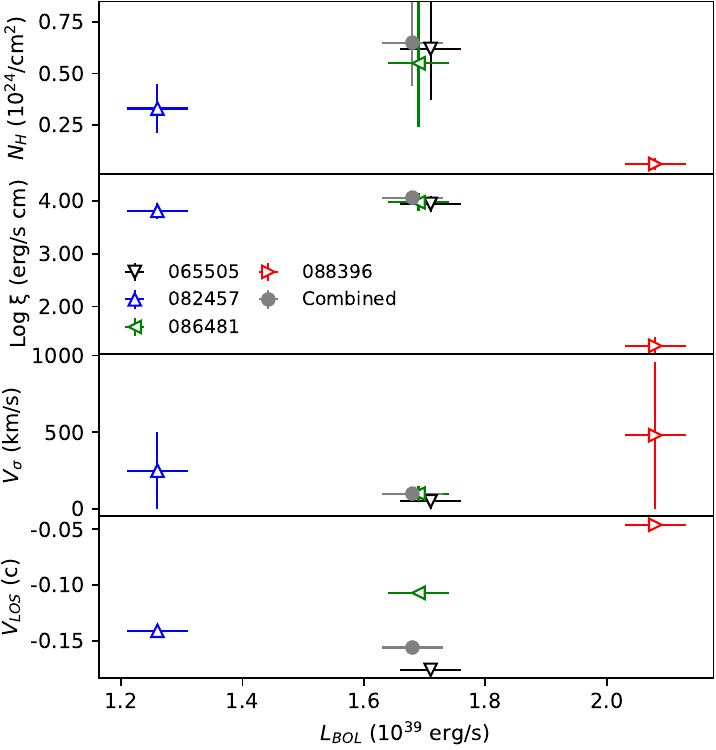}
   \vspace{-0.1cm}
   \caption{Summary plot for the results relative to the \texttt{xabs} (photoionised absorber) model. The grey circles refer to the combined spectrum.}
              \label{fig:results_summary_xabs}
   \vspace{-0.1cm}
    \end{figure}

\begin{figure}
   \centering
   \includegraphics[scale=0.7]{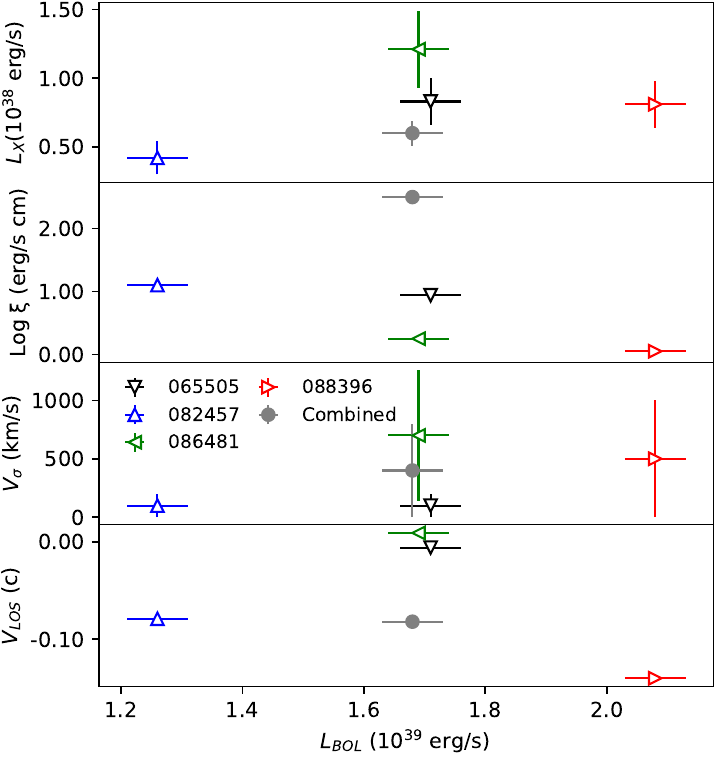}
   \vspace{-0.1cm}
   \caption{Summary plot for the results relative to the \texttt{pion} (photoionised emitter) model. The grey circles refer to the combined spectrum.}
              \label{fig:results_summary_pion}
   \vspace{-0.1cm}
    \end{figure}

\begin{figure}
   \centering
   \includegraphics[scale=0.7]{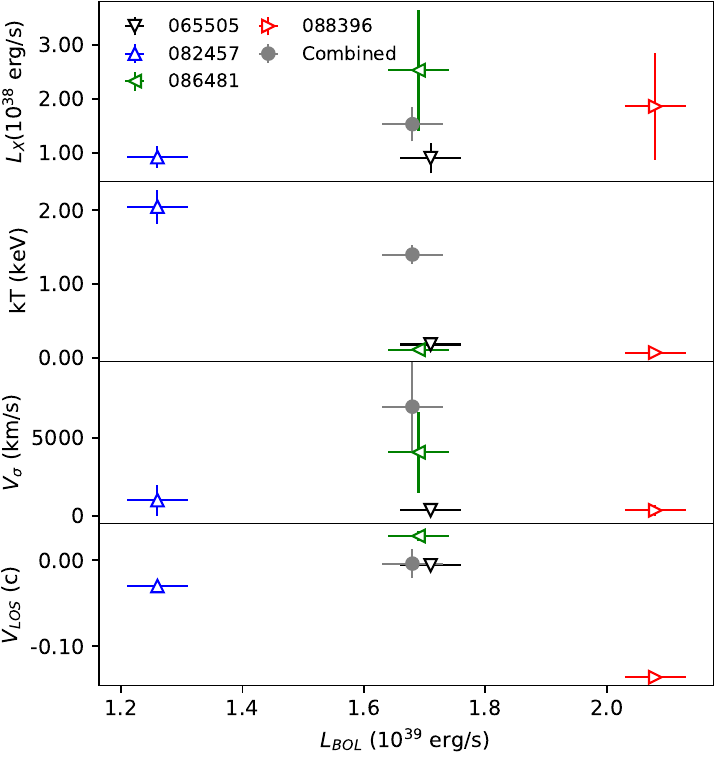}
   \vspace{-0.1cm}
   \caption{Summary plot for the results relative to the \texttt{cie} (collisionally-ionised emitter) model. The grey circles refer to the combined spectrum.}
              \label{fig:results_summary_cie}
   \vspace{-0.1cm}
    \end{figure}

    In Figs.\,\ref{fig:results_summary_xabs}, \ref{fig:results_summary_pion} and \ref{fig:results_summary_cie} we compare the best-fit values of the main parameters of the emitting and absorbing plasmas obtained for individual observations and (in red) for the time-averaged spectrum. We remind the reader that two of the four observations (obsid 065505 and 086481) occurred at similar luminosities, whereas the other two occurred during low-flux and flaring periods. The two equi-luminosity observations that occurred at the most common flux regimes of NGC 55 ULX-1 exhibit consistent wind properties with a dominant low-velocity emitting plasma and a mildly-relativistic outflow {($|v_{\rm LOS}|\gtrsim0.15c$)}. The luminosity corrections are comparable and amount to 60-70\%. The RMS are also consistent between these two observations (12-15\%).
       
    The low flux observation (obsid 082457) shows similar wind properties for the absorber, indicating that the source is still in a super-Eddington regime. The spectra hardness, HR, is broadly constant between the different observations, indicating a similar state. However, the lower RMS and luminosity correction suggest that the source is intrinsically fainter during obsid 082457, which is confirmed by the lower significance and two-times fainter line-emitting component (see Table\,\ref{table:wind_properties}).
    
    The highest-flux observation (obsid 088396) caught the tail of a major flare. At the peak of the flare the observed flux was twice higher than the average level; see Fig.\,\ref{fig:swift}. Obsid 088396 is characterised by a moderate RMS (8\%) and luminosity correction, indicating that some obscuration might still be present. Here, the fastest {($-0.1c$)} absorbing component seen in the other observations appears clearly weakened while there is a potential $3\,\sigma$ detection of an emerging component with both lower velocity {($v_{\rm LOS}=-0.05c$)} and ionisation parameter ($\log\xi=1.3$). This situation is remarkably similar to what was observed during the high-flux regime of NGC 1313 ULX-1, which was interpreted as due to the extension of the launching radius or spherisation radius, defined as $R_{\rm sph}=R_{\rm in} \dot{M}/\dot{M}_{\rm Edd}$ with $R_{\rm in} = 6 \, R_{\rm G}$, at increasingly higher super-Eddington rates \citep{Pinto2020b}. The fastest component might have weakened due to an increase of its ionisation parameter (which is proportional to the luminosity since $\xi=L_{\rm ion} / n_{\rm H} R^2$) similarly to disappearing winds in highly-accreting supermassive black holes (see, e.g., \citealt{Pinto2018a}). An intrinsically higher luminosity might be the reason behind the appearance of a fast line-emitting plasma with the exceptionally high Doppler shift {($-0.14c$)}.

    \subsection{ULX feedback and bubble nature}
    The kinetic power of the wind is defined as $L_w = 1/2 \, \dot{M}_w \, v_w^2$, where $\dot{M}_w = 4 \, \pi \, R^2 \, \rho \, v_w \, \Omega \, C$ is the outflow rate, $\Omega$ and $C$ are the solid angle and the volume filling factor (or \textit{clumpiness}). As previously done for other ULXs, we may adopt $\Omega$ and $C\sim0.1$ as conservatively determined from theoretical simulations of winds driven by radiation pressure in super-Eddington winds \citep{Takeuchi2013,Kobayashi2018}. We can then resolve for the density $\rho=n_{\rm H} \, m_p \, \mu$ (with $m_p$ the proton mass and $\mu = 0.6$ the average particle weight of a highly ionised plasma) and the distance $R$ from the ionising source. Using the definition of $\xi$ {(under the assumption that the wind remains constant which might not be true although the higher significance in the combined spectrum would suggest it)} we can get rid of the $R^2 \rho$ factor. The wind power is estimated to be around $10^{40} \, {\rm erg/s}$; taking an even more conservative approach (see, e.g., Eq.\,(23) in \citealt{Kobayashi2018}), we would obtain a wind power $L_w > 10^{39} \, {\rm erg/s}$ (i.e. $>30$\,\% of the total budget if compared to the luminosity). This is sufficient to power the typical interstellar bubbles found around many ULXs including NGC 55 ULX-1 (see, e.g., \citealt{Gurpide2022,Zhou2023}).
    
    \subsection{NewAthena and the access to short timescales}
\begin{figure}
   \centering
   \includegraphics[scale=0.33]{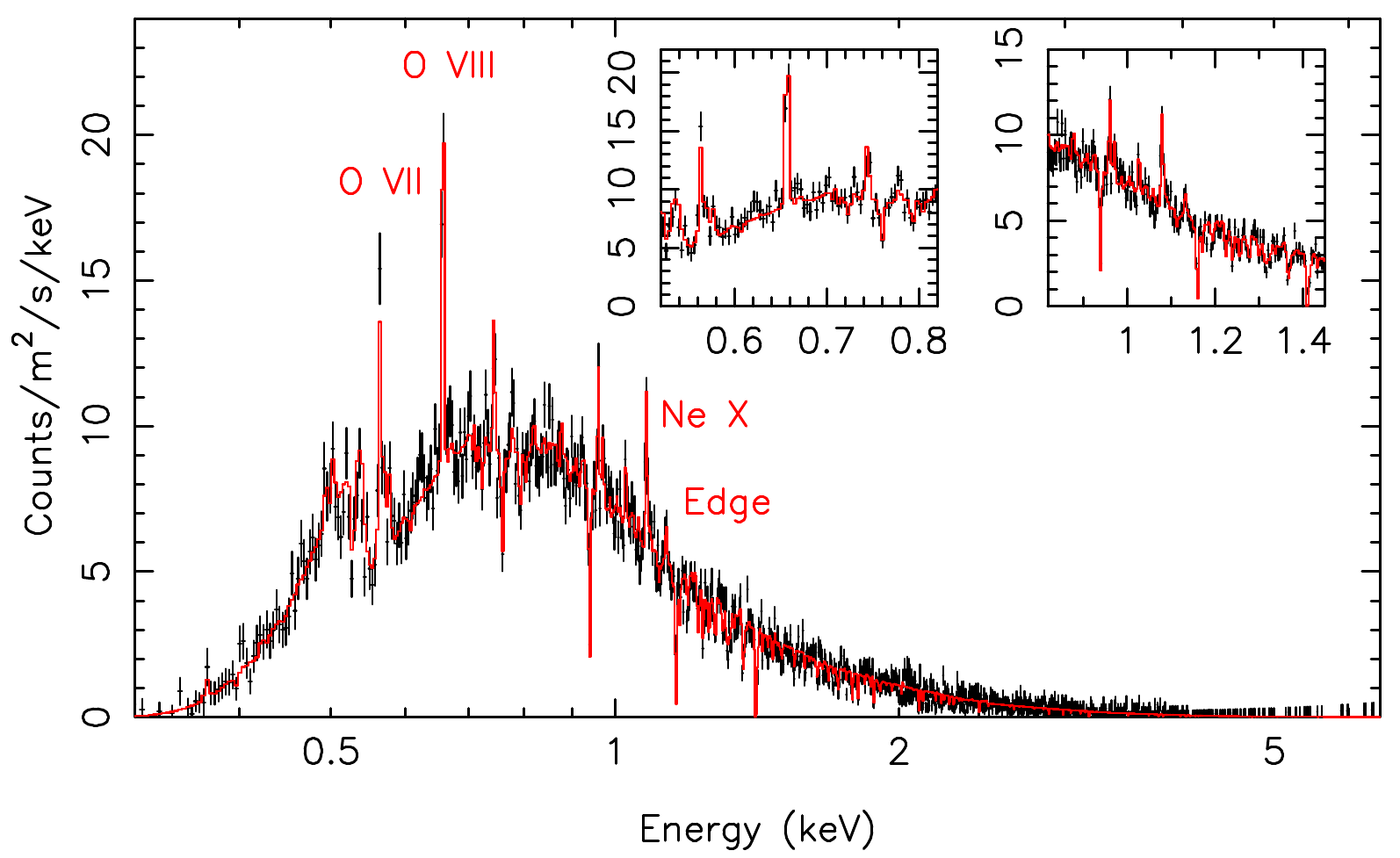}
   \vspace{-0.1cm}
   \caption{Simulation with NewAthena / X-IFU (10 ks). Template model used {(red line)}: best-fit continuum + wind model from obsid 0655050101 (see Table\,\ref{table:wind_properties} and Fig.\,\ref{fig:spec_bestfit}, top-left panel).}
              \label{fig:new_missions}
   \vspace{-0.1cm}
    \end{figure}

    Despite our XMM-Newton campaigns {having unveiled many interesting properties of nearby ULXs, crucial details in the role of the wind in the spectral transitions and the geometry of the accretion flow are still missing.} We have shown that observations at different epochs and fluxes are key to answering some questions, but we are currently limited on the timescales that can be explored in great detail (generally above 100\,ks). Flux dips and flares can be as short as a few ks in both hard and soft ultra-luminous sources (see, e.g., \citealt{Alston2021,DAi2021, Pintore2025} {and B22 for NGC 55 ULX-1}). In the future, the X-IFU microcalorimeter onboard NewAthena will enable exploration of this phenomenology thanks to its unique combination of high effective area, spectral resolution, and bandpass \citep{Peille2025}. To showcase the capabilities of NewAthena, we performed a simulation adopting the reformulated (2024) response matrix of X-IFU assuming an exposure time of 10 ks and as template model the best-fit continuum + wind model obtained for obsid 0655050101 (see Table\,\ref{table:wind_properties} and Fig.\,\ref{fig:spec_bestfit}, top-left panel). The simulated spectrum exhibits a plethora of emission and absorption lines, many of which were individually detected at confidence levels well above $5\,\sigma$ (see Fig.\,\ref{fig:new_missions}). This confirms that we will be able to track down the wind changes on timescales of the continuum variability, which is necessary to search for short-term responses and mutual dependence between accretion and ejection properties.

\section{Conclusions}\label{sec:conclusions}

In this work, we have undertaken a deep observing campaign with XMM-Newton of a well-known ultra-luminous X-ray source, NGC 55 ULX-1, in order to 1) corroborate the previous evidence of outflows in the high-resolution RGS spectra and 2) to search for any trend between the accretion (e.g. luminosity, spectral hardness and variability) and ejection (wind velocity, column density and temperature) properties. The combined time-average spectrum strengthened the previous detection confidence levels and provided more insight on the nature of the 1\,keV feature usually seen in high-count CCD spectra of ULXs. The comparison between results obtained at different epochs revealed that the wind responds to the variability of the underlying continuum, and these variations may be used to understand the actual accretion regime and the nature of the source. Future observations with high-throughput / spectral-resolution facilities, such as NewAthena, will cement our understanding thanks to the possibility of tracing such trends on the short variability timescales of the accretion process.

\section*{Data Availability}

All data and software used in this work are publicly available from the XMM-Newton Science Archives (\url{https://www.cosmos.esa.int/web/xmm-newton/xsa}), (\url{https://heasarc.gsfc.nasa.gov/}). Our codes are also publicly available on GitHub (\url{https://github.com/ciropinto1982}).

\begin{acknowledgements}
We acknowledge funding from PRIN MUR 2022 SEAWIND 2022Y2T94C, supported by European Union - Next Generation EU, Mission 4 Component 1 CUP C53D23001330006, INAF Large Grant 2023 BLOSSOM F.O. 1.05.23.01.13 and  the Italian Space Agency, contract ASI-INAF program I/004/11/4-6. The authors also thank Srimanta Banerjee for helpful discussions on accretion disc physics.
\end{acknowledgements}

\bibliographystyle{aa}
\bibliography{ref.bib}

\appendix
    
\section{Technical details}
\label{sec:appendix}

In this section, we include some technical details and plots that were excluded from the main body of the paper.

\subsection{SED and plasma thermal stability}
\label{appendix:thermal_stability}

\begin{figure}
   \centering
   \includegraphics[scale=0.55]{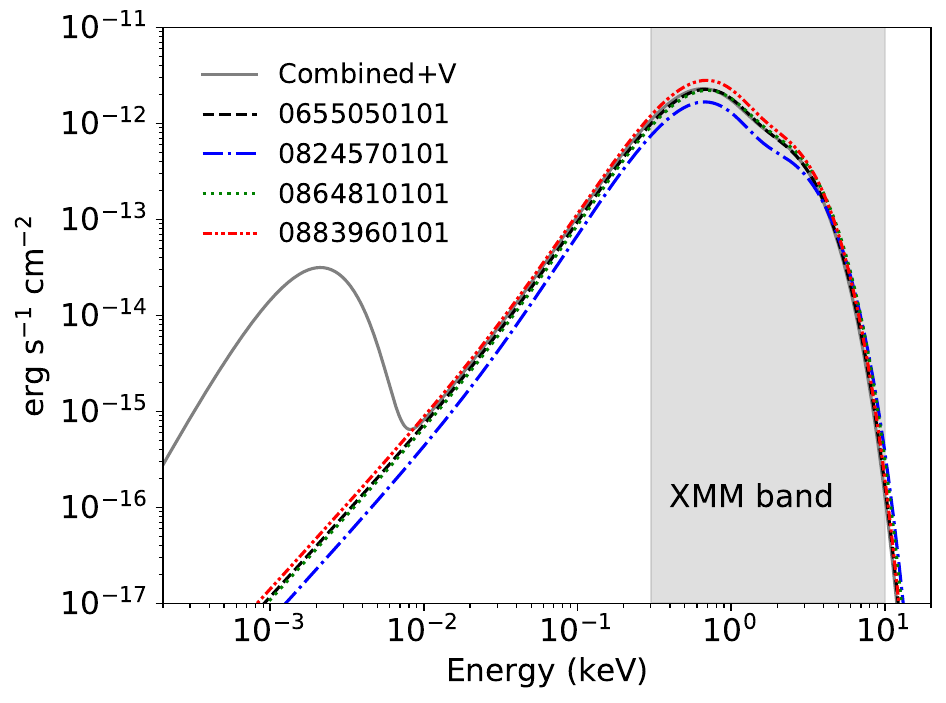}
   \includegraphics[scale=0.5]{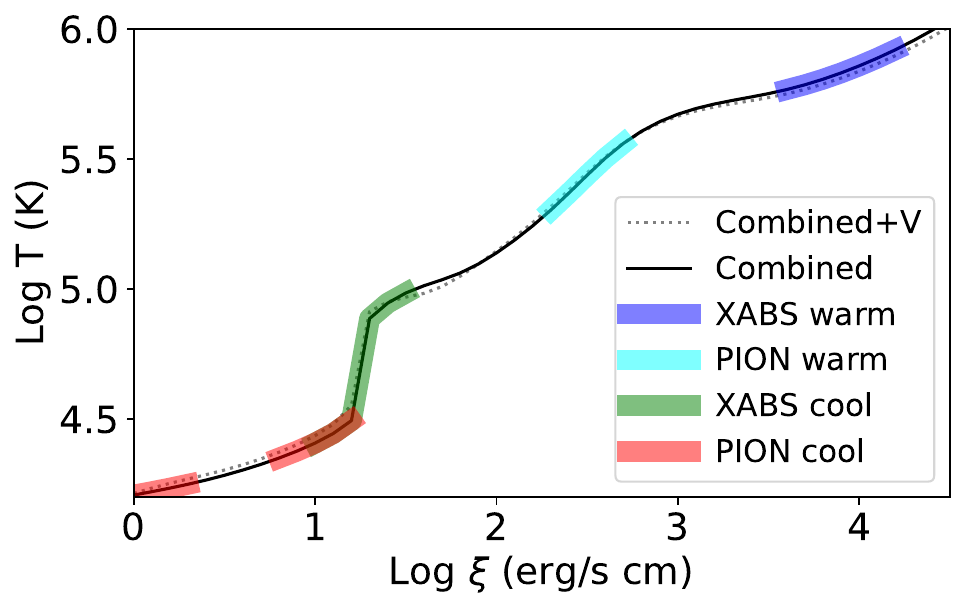}
   \includegraphics[scale=0.5]{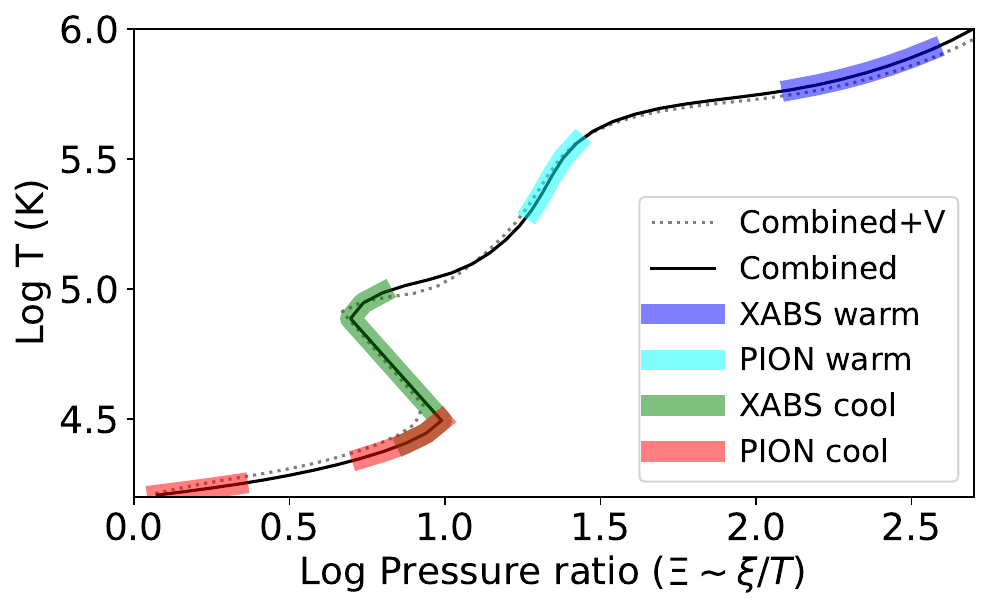}
\vspace{-0.1cm}
   \caption{Spectral energy distribution (top panel), {ionisation balance (middle panel)} and stability curve (bottom panel). The band covered by XMM-Newton is shown in grey. With ``Combined+V" we refer to the multi-wavelength SED built with XMM-Newton X-ray data and HST optical data. The {ionisation balance and} stability curves of the individual observations are not shown since they are almost superimposable. The best-fit ranges of the emission (\texttt{pion}) and absorption (\texttt{xabs}) line components as obtained for the individual observations are shown as thicker segments (see  Table\,\ref{table:wind_properties}).}
    \label{fig:sed_ionbal}
\vspace{-0.1cm}
\end{figure}

{As mentioned in Sect.\,\ref{sec:data_reduction}, no optical counterpart was detected in the XMM-Newton/OM data. However, we made sure that the exclusion of the optical/UV emission in our SED model has a negligible impact on the ionisation balance. As can be seen in Fig.\,\ref{fig:sed_ionbal}, the extrapolation of our best-fit model down to 0.002 keV results in a flux of $5\times10^{-17}$ erg/s/cm$^2$, which is a factor 10x lower than that measured with MUSE (\citealt{Zhou2023}) and even lower than previous estimates with HST which showed a very faint optical counterpart beyond the reach of OM (HST magnitude $\gtrsim23$, \citealt{Gladstone2013}).}

For a comparison, we have also built a multi-wavelength SED (and computing the corresponding ionisation balance) by adding the HST optical data (see, e.g., \citealt{Gladstone2013,Zhou2023}).
{For this task, we adopted a plasma density $n_{\rm H} = 10^{10}$ cm$^{-3}$ in coherence with the assumptions for photoiosation model scans which were suggested by the independent gaussian modeling because the balance depends on the density (see, e.g., \citealt{Bianchi2017}).}
There is no significant difference in the stability curves due to the weakness of the flux at optical-UV wavelengths (see Fig.\,\ref{fig:sed_ionbal} dotted line) and the rather soft SED, which is a characteristic common to ultra-luminous X-ray sources (see \citealt{Pinto2023a} and references therein).

{In Fig.\,\ref{fig:sed_ionbal}, we have also highlighted, with thicker segments, the best fits of photoionised plasma models with emission lines (\texttt{pion}) and absorption lines (\texttt{xabs}) from Table\,\ref{table:wind_properties} to check whether all solutions correspond to stable branches of the stability curves. This occurs for most components, with the exception of the {slow/cool ($v_{\rm LOS}=-0.05c$ and $\log\,\xi=1.26$)} absorbing plasma that appears in the highest-flux obsid (088396) {and, at a lower-significance, in the first observation}. The solutions are labelled as \textit{warm/cool} to differentiate between the high/low $\xi$ values in Table\,\ref{table:wind_properties}.}

\subsection{Spectra {and fits} of individual observations}
\label{appendix:individual_fits}

\begin{figure*}
   \centering
   \includegraphics[scale=0.330500]{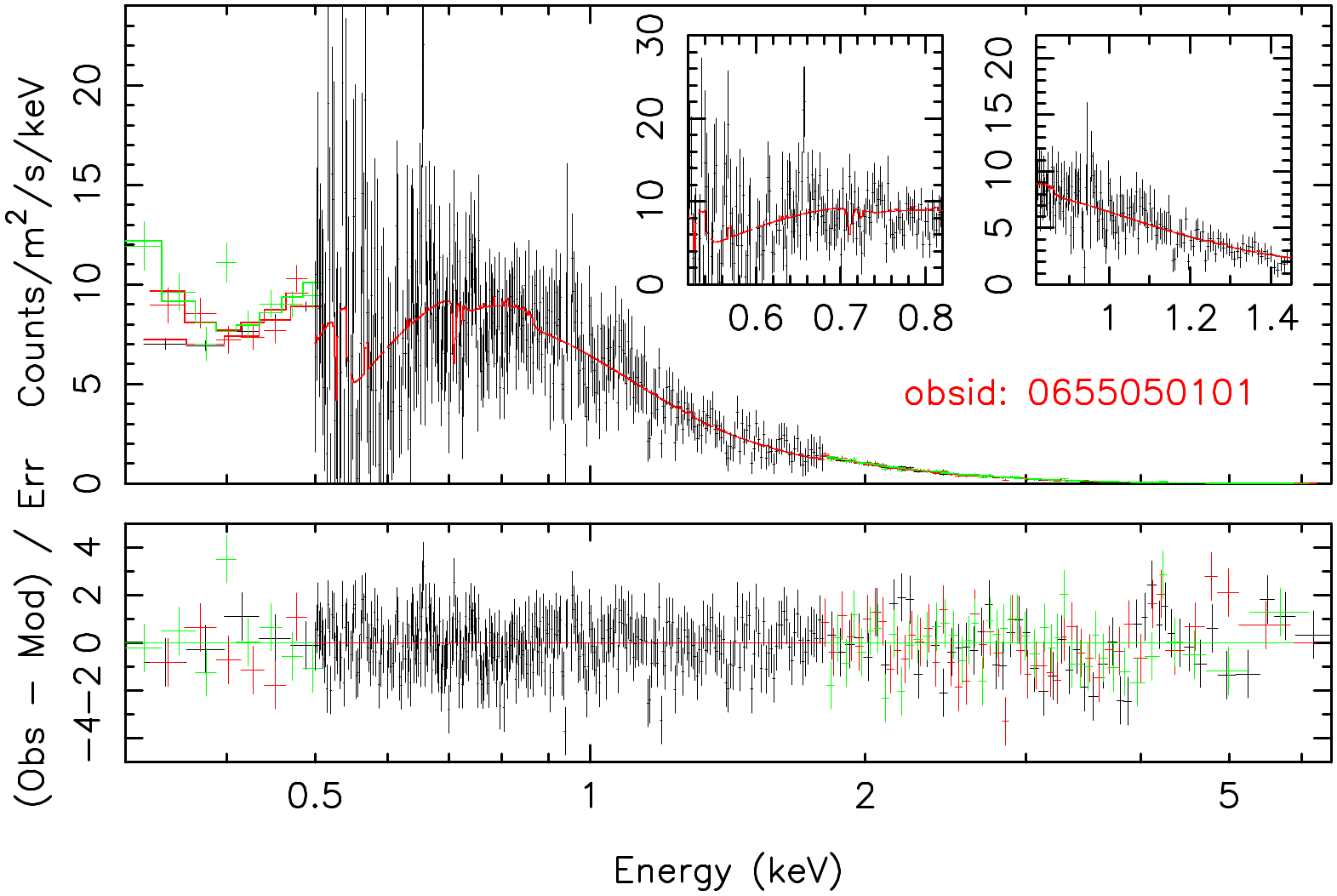} \hspace{0.5cm}
   \includegraphics[scale=0.330500]{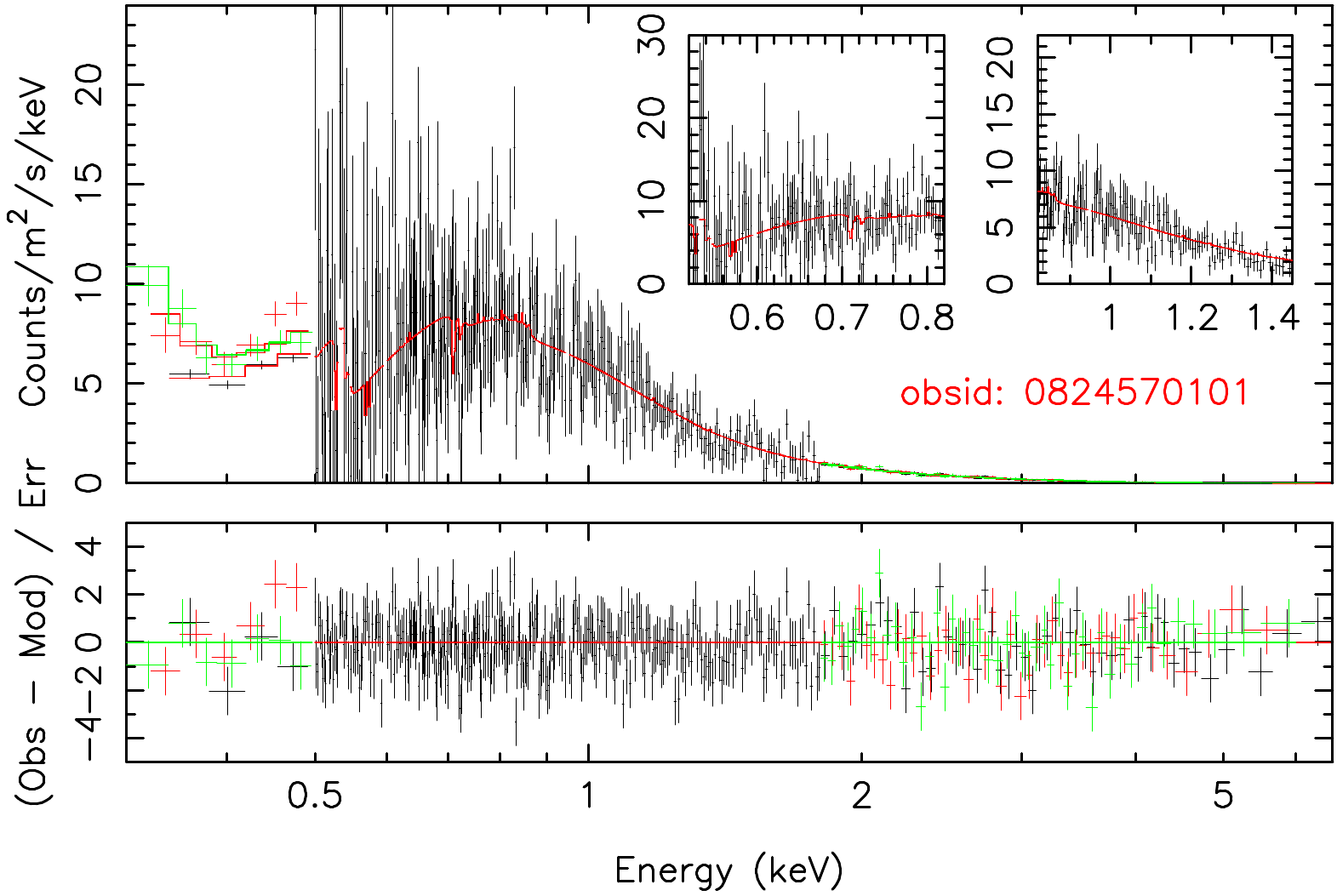}
   \includegraphics[scale=0.330500]{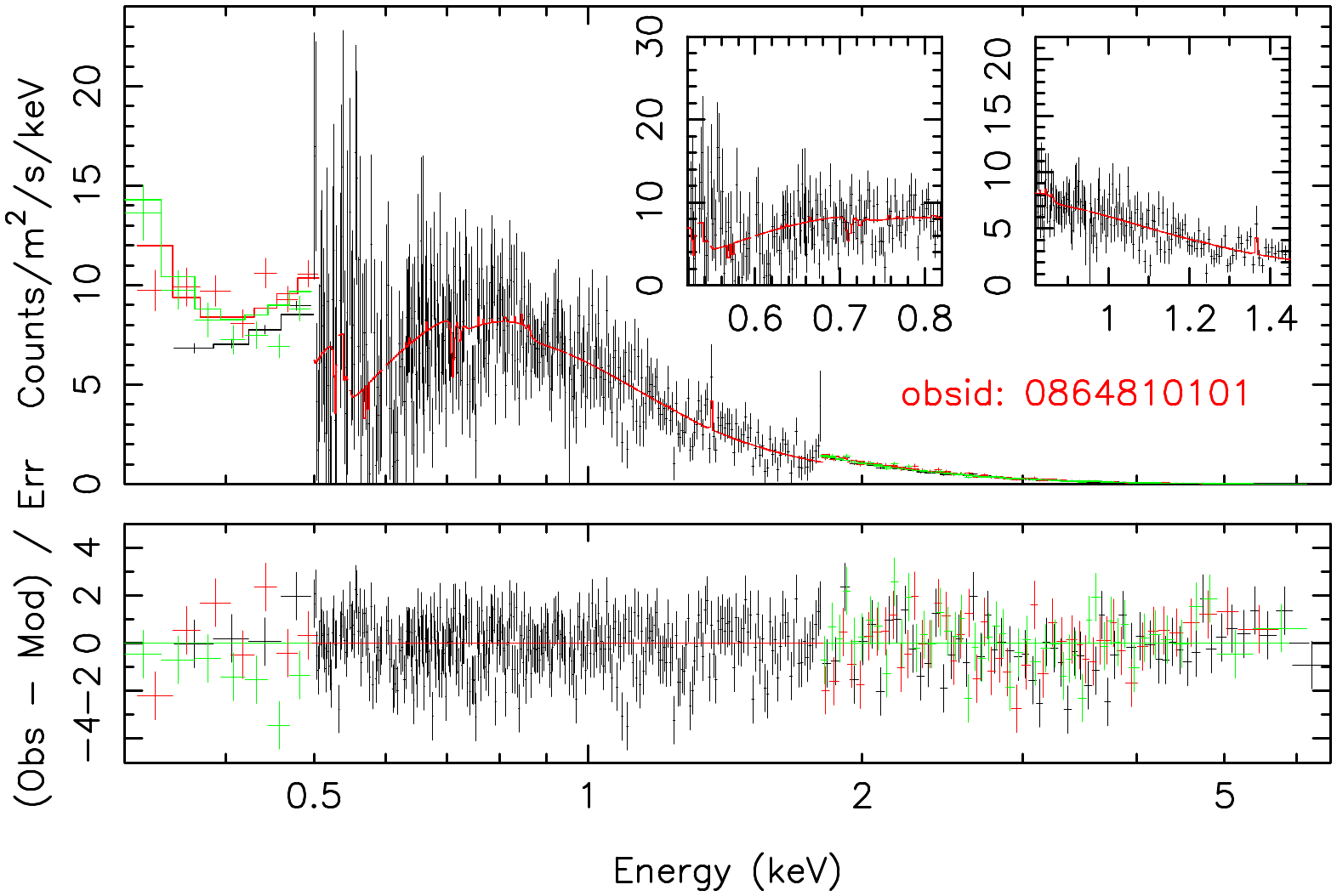} \hspace{0.5cm}
   \includegraphics[scale=0.330500]{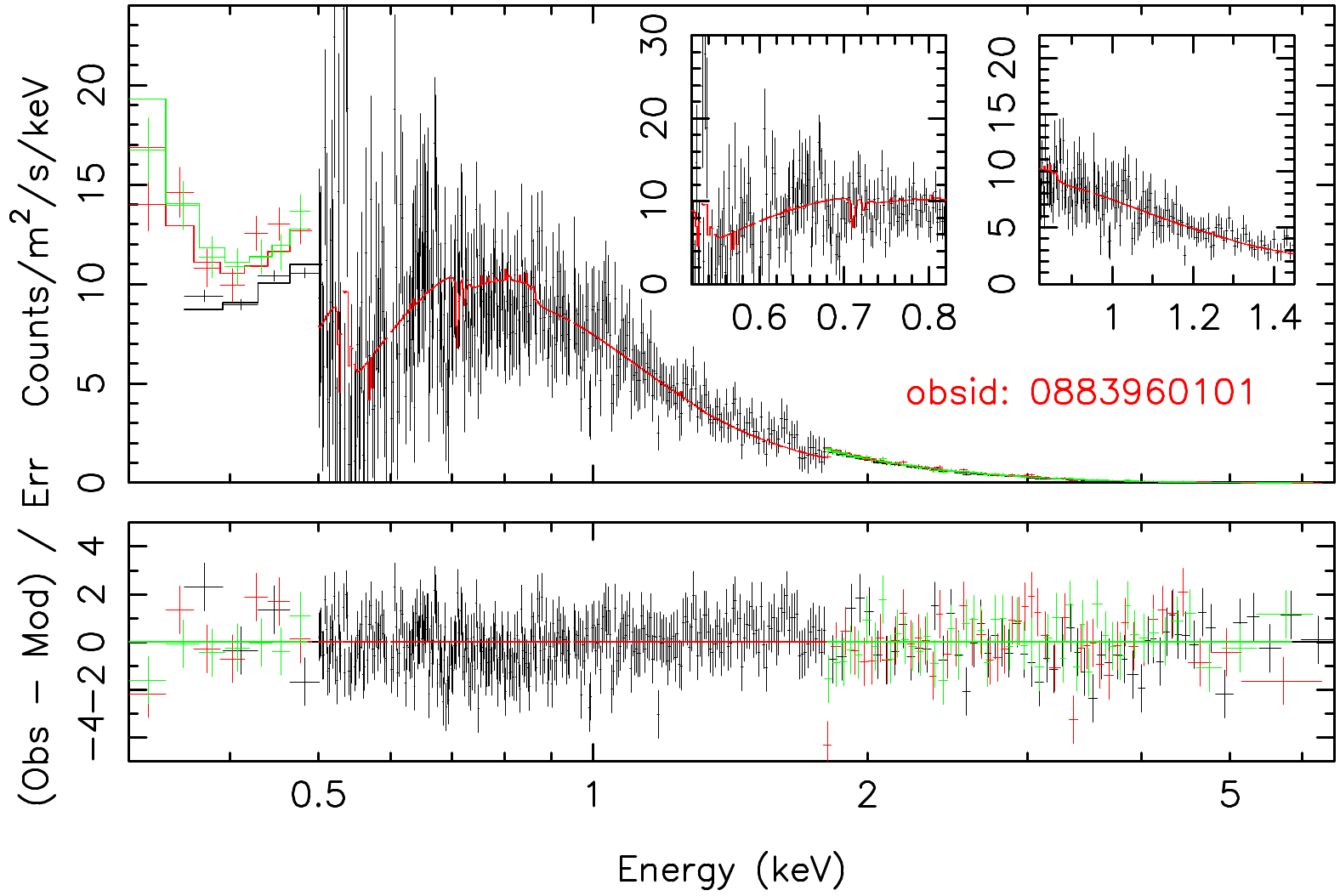}
   \caption{XMM-Newton spectra and continuum modelling of the individual observations. Labels are same as in Fig.\,\ref{fig:spec_tavg}. Only the 1st order RGS spectrum is used due to the low statistics of the 2nd order individual spectra.}
              \label{fig:spec_continuum}
\vspace{-0.1cm}
    \end{figure*}

\begin{figure*}
   \centering
   \includegraphics[scale=0.330500]{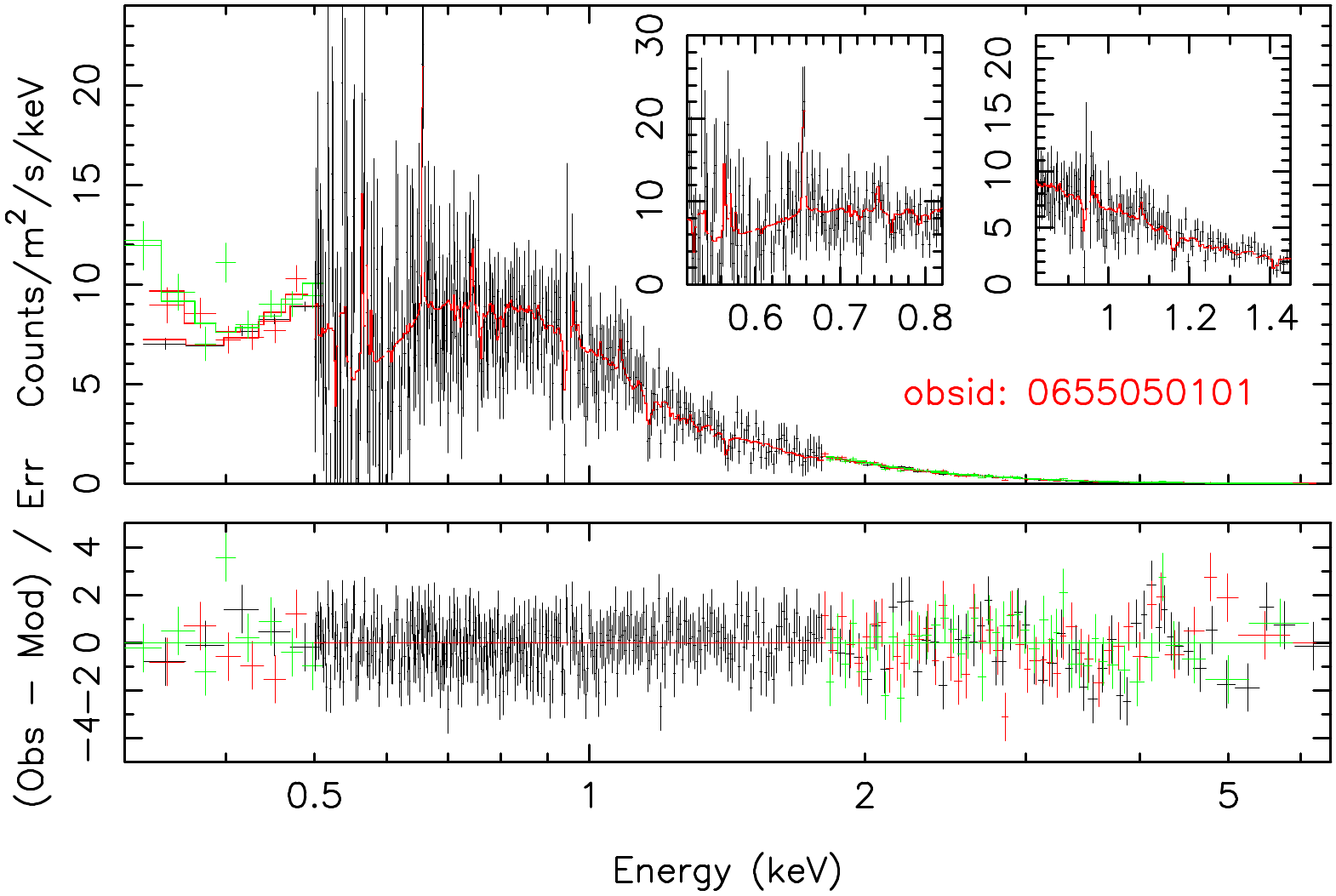} \hspace{0.5cm}
   \includegraphics[scale=0.330500]{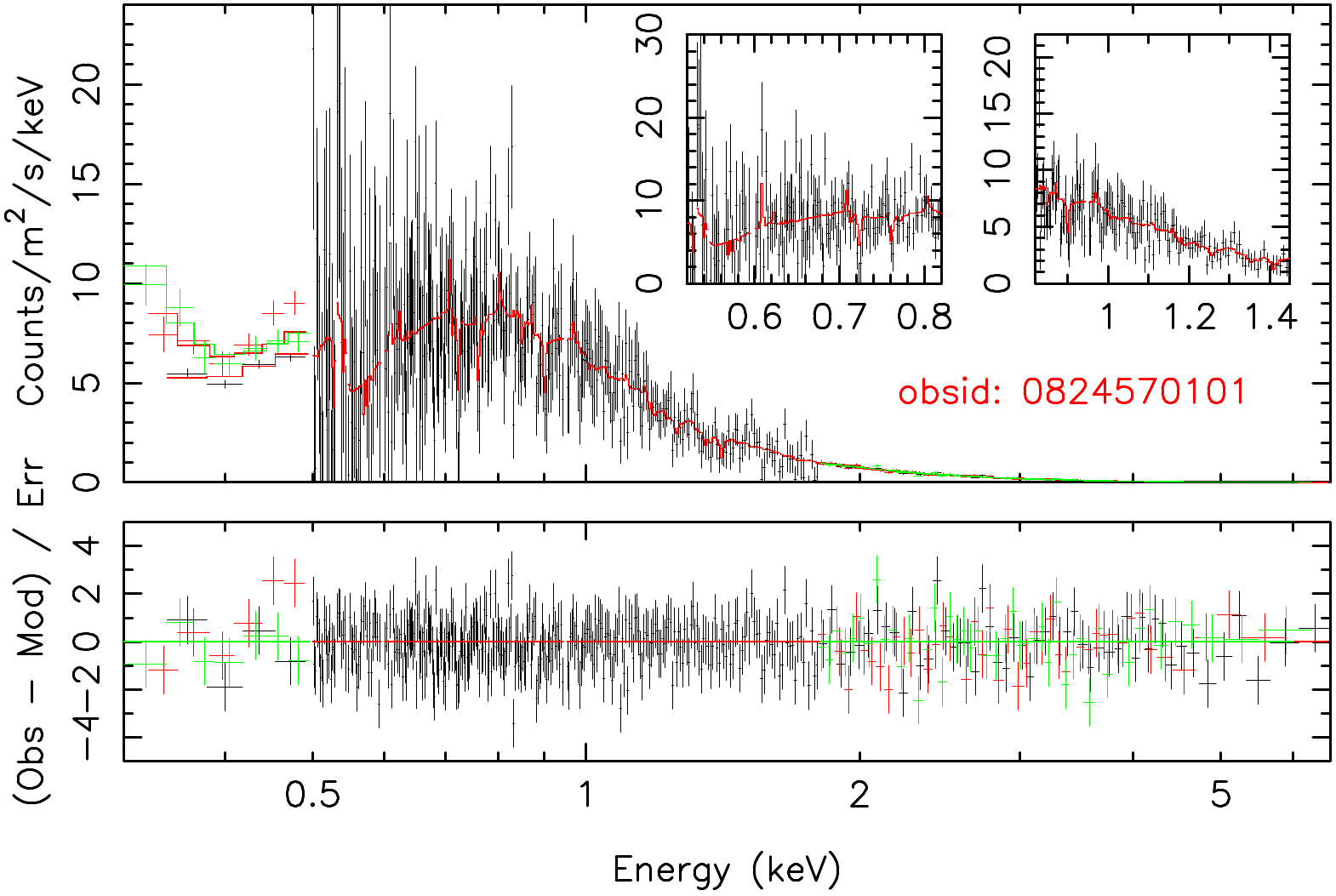}
   \includegraphics[scale=0.330500]{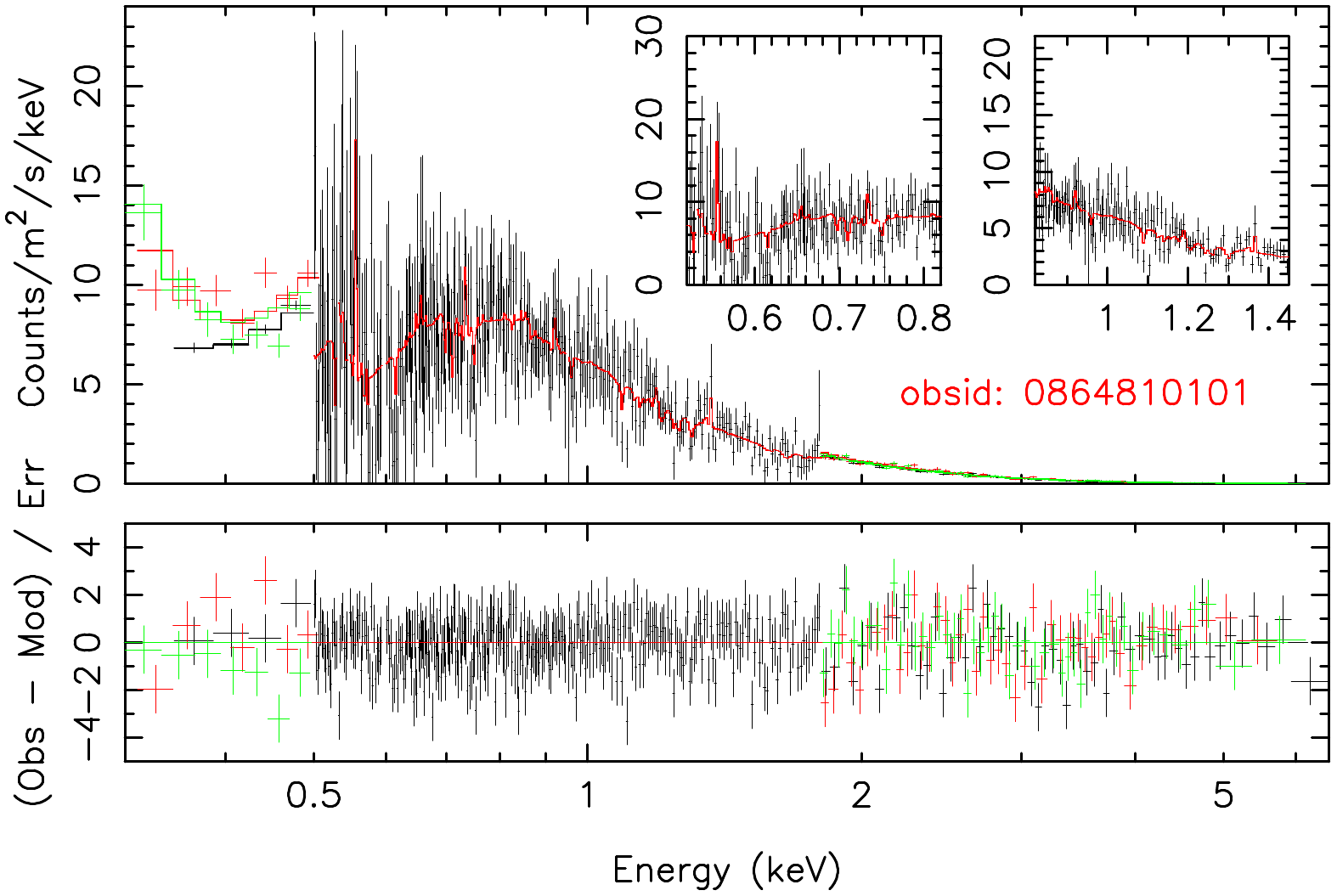} \hspace{0.5cm}
   \includegraphics[scale=0.330500]{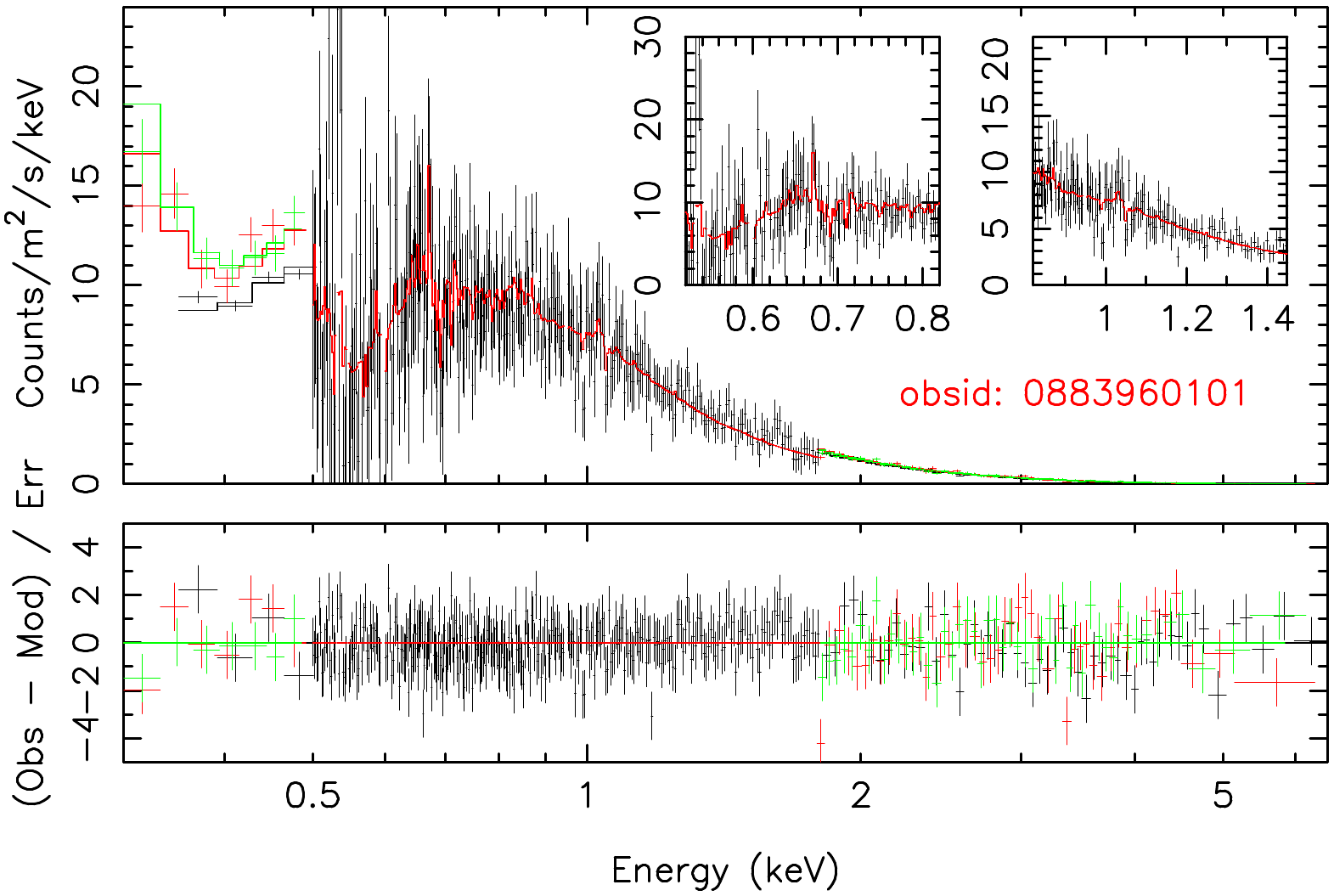}
   \caption{XMM-Newton spectra and best-fit photoionised emission/absorption model of the individual observations. Labels are same as in Fig.\,\ref{fig:spec_continuum}.}
              \label{fig:spec_bestfit}
\end{figure*}

{In Fig.\,\ref{fig:spec_continuum}, we show the spectra of each observation with the spectral continuum model overlaid. We adopted a double thermal model consisting of a cool (0.16--0.21 keV) blackbody and a warmer (0.7-0.9 keV) blackbody modified by coherent Compton scattering for the broader and harder component. These two components refer to the outer/cooler and inner/hotter super-Eddington inner disc, with the latter being slightly obscured due to the LOS inclination. More detail is provided in Sect.\,\ref{sec:continuum_and_SED}. In Fig.\,\ref{fig:spec_bestfit}, we show the best-fit continuum modelling of the individual observations with the addition of two photoionised plasma components to account for the emission (\texttt{pion}) and the absorption (\texttt{xabs}) lines. For more detail, see Sect.\,\ref{sec:individual_obs}.}

\begin{figure*}
   \centering
   \includegraphics[scale=0.325]{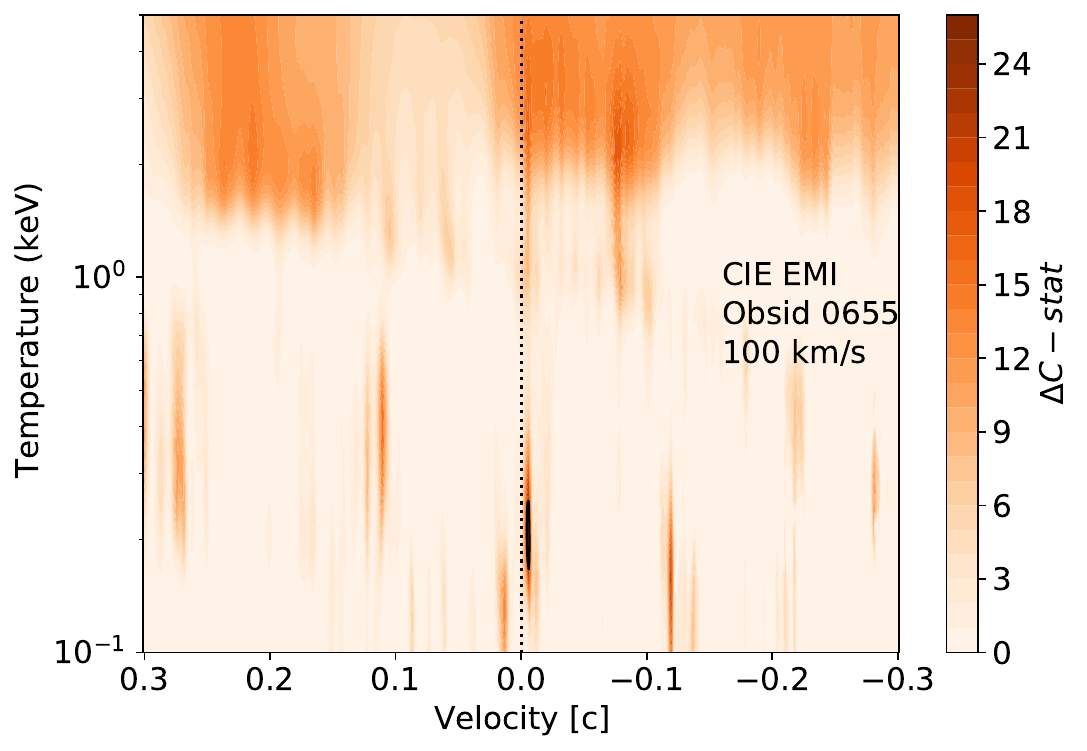}
   \includegraphics[scale=0.325]{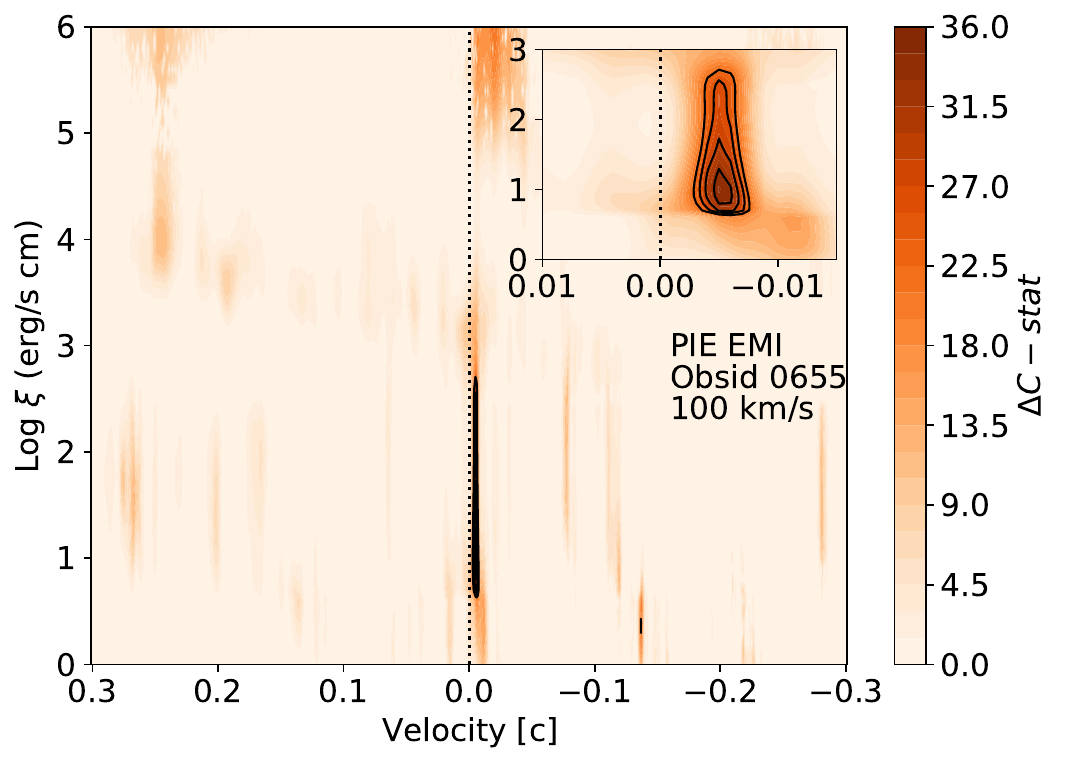}
   \includegraphics[scale=0.325]{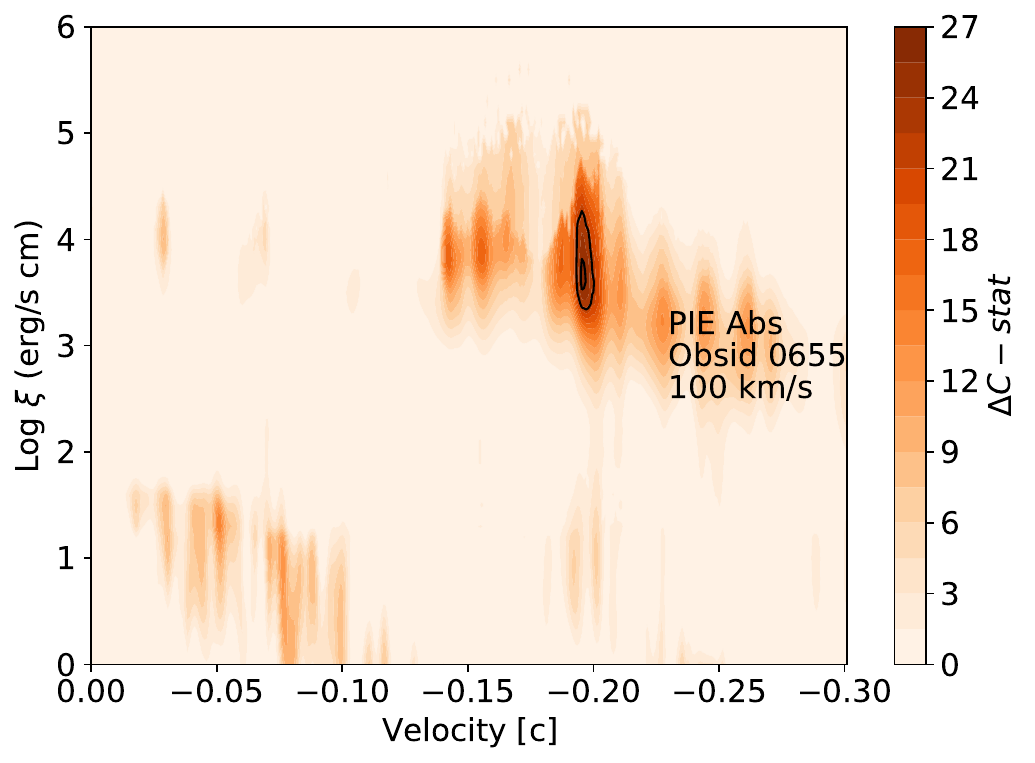}
   \includegraphics[scale=0.325]{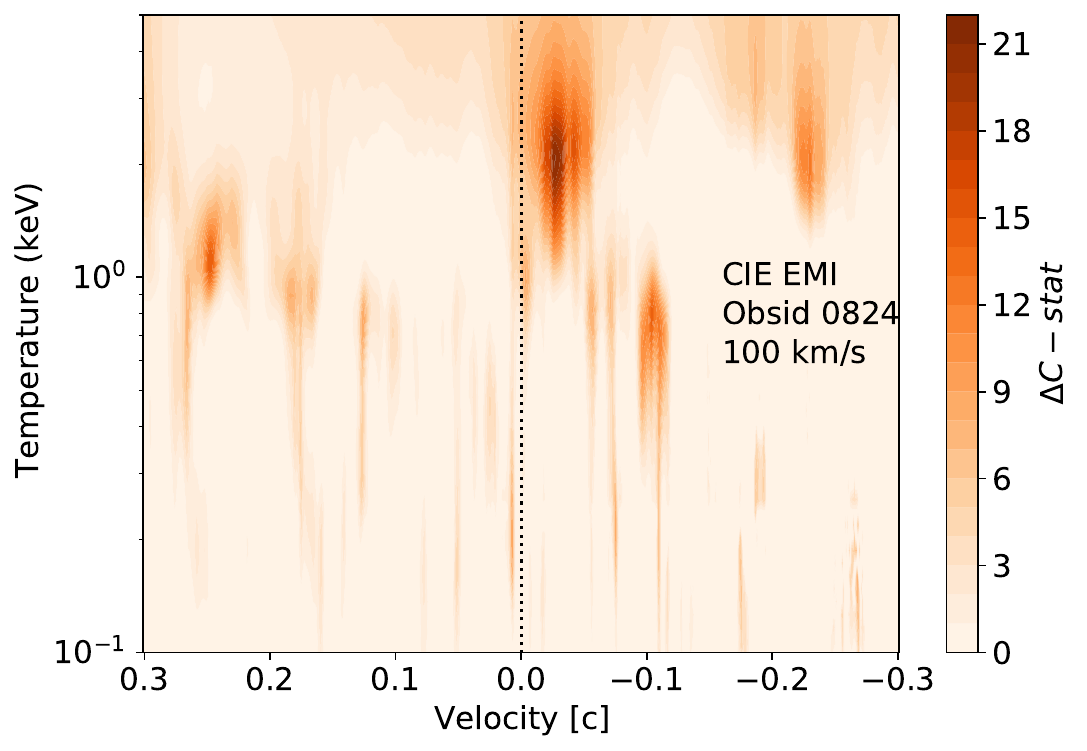}
   \includegraphics[scale=0.325]{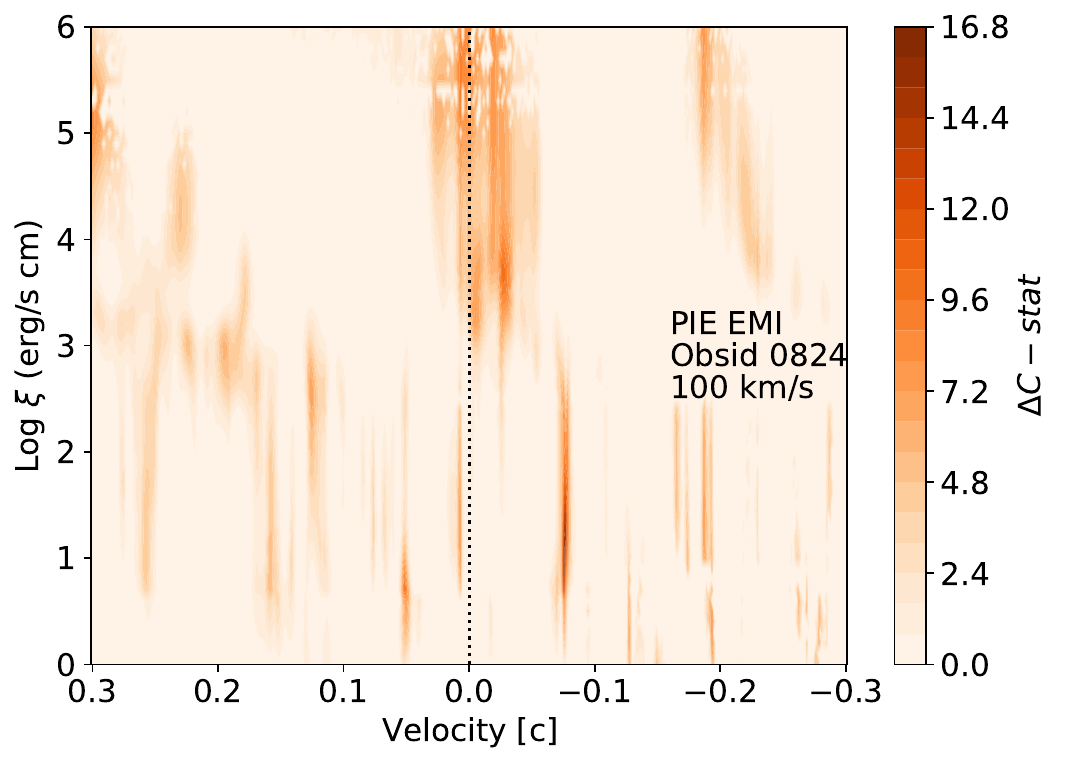}
   \includegraphics[scale=0.325]{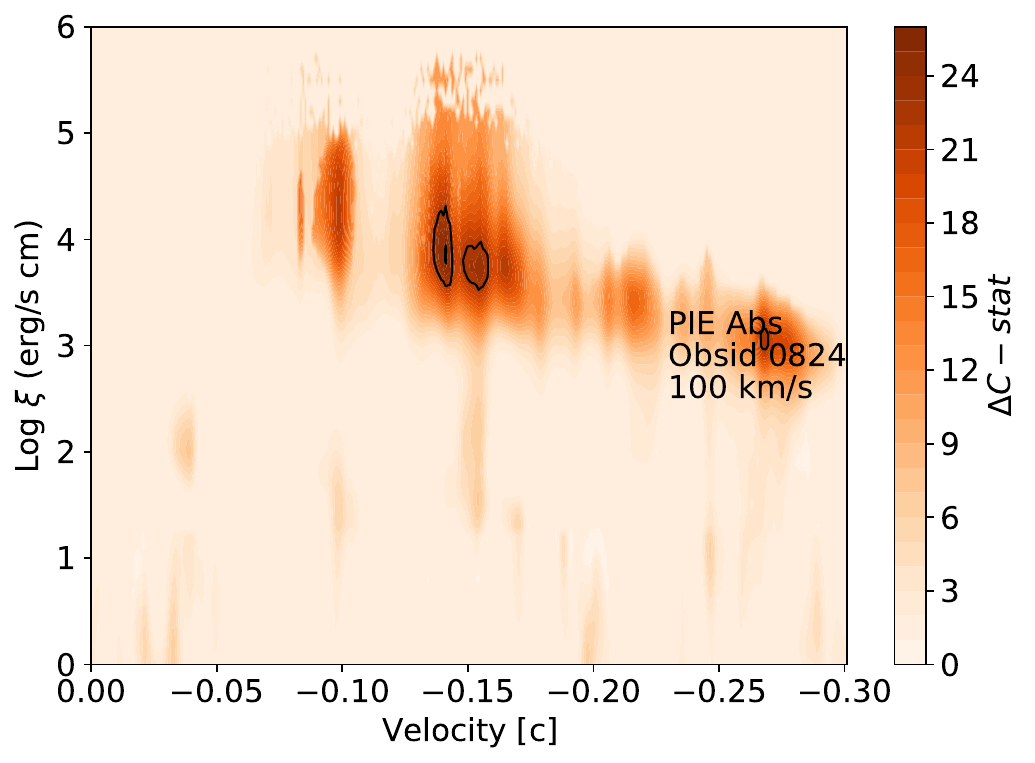}
   \includegraphics[scale=0.325]{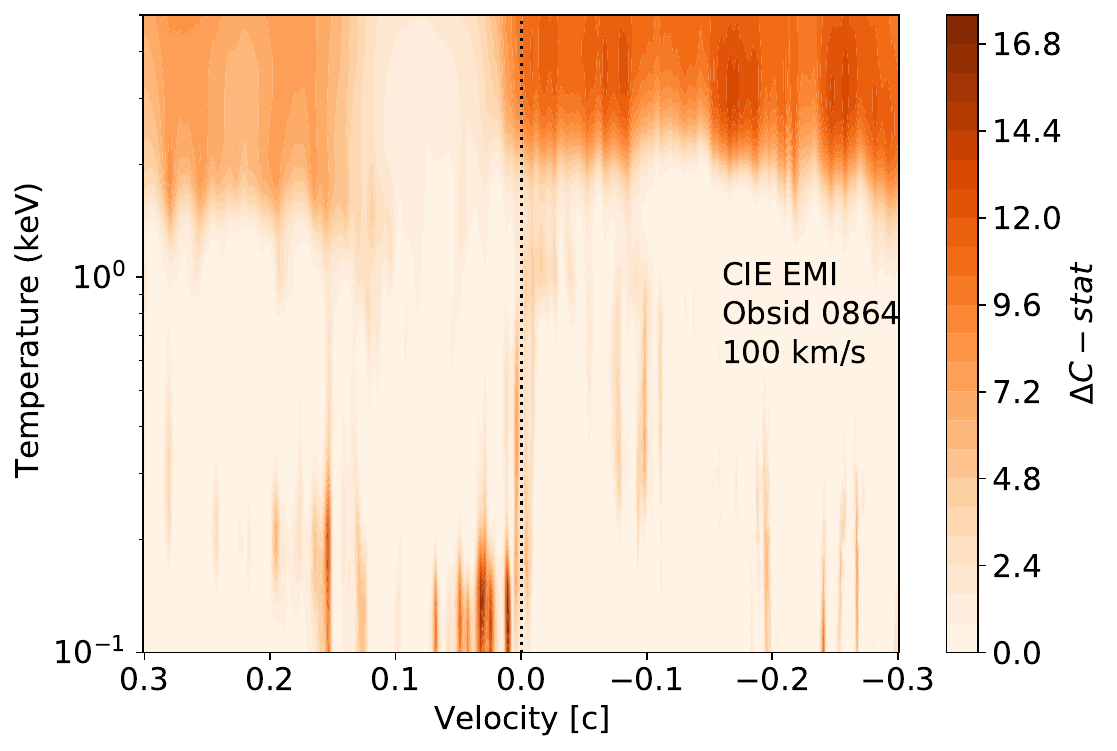}
   \includegraphics[scale=0.325]{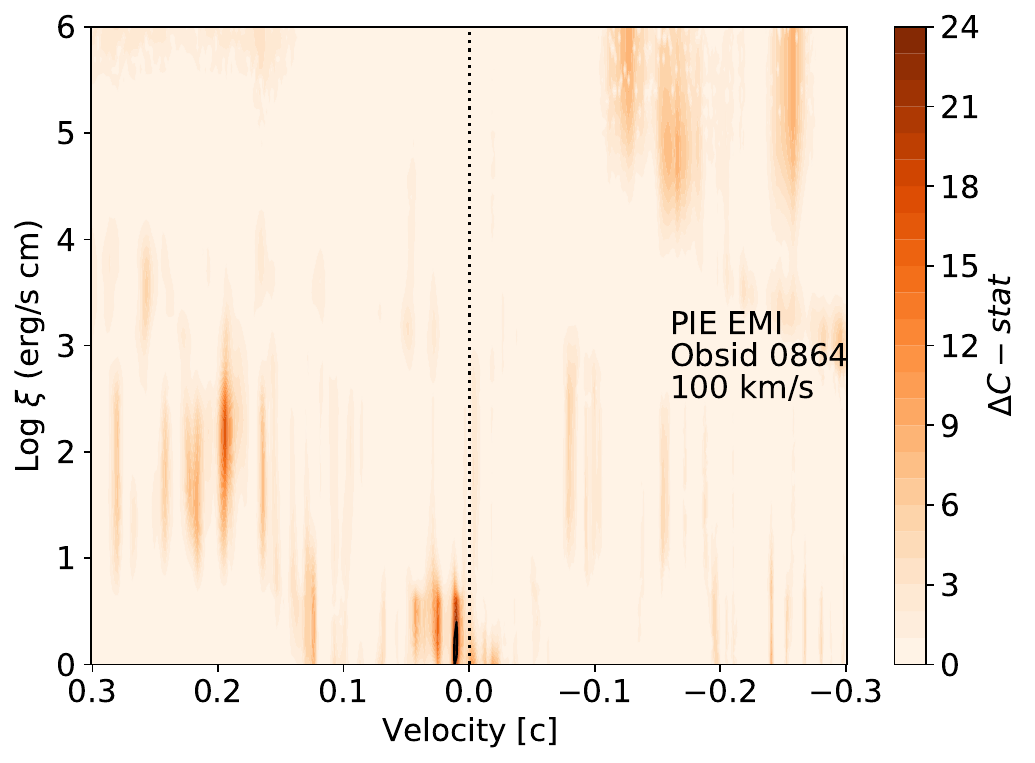}
   \includegraphics[scale=0.325]{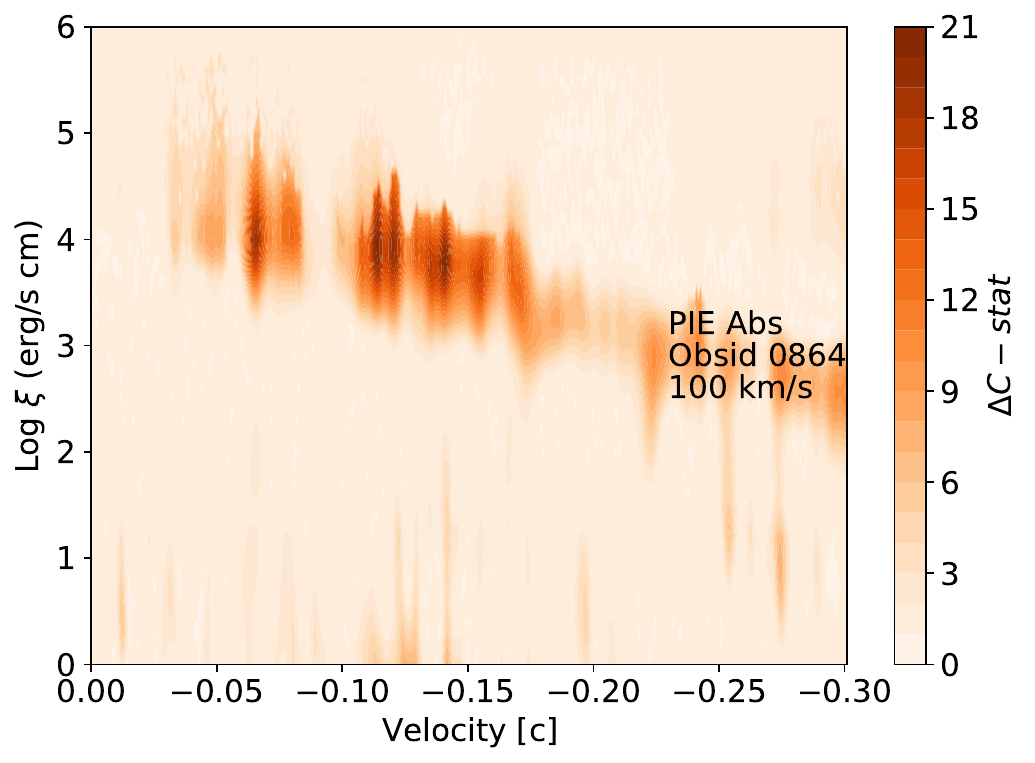}
   \includegraphics[scale=0.325]{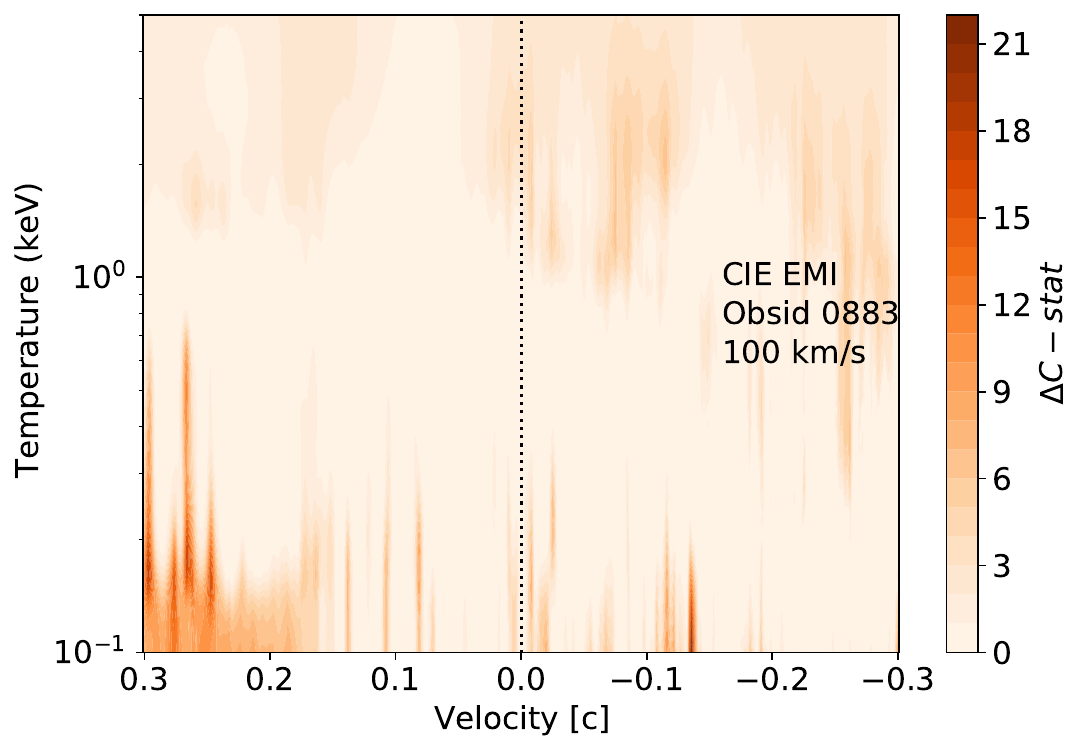}
   \includegraphics[scale=0.325]{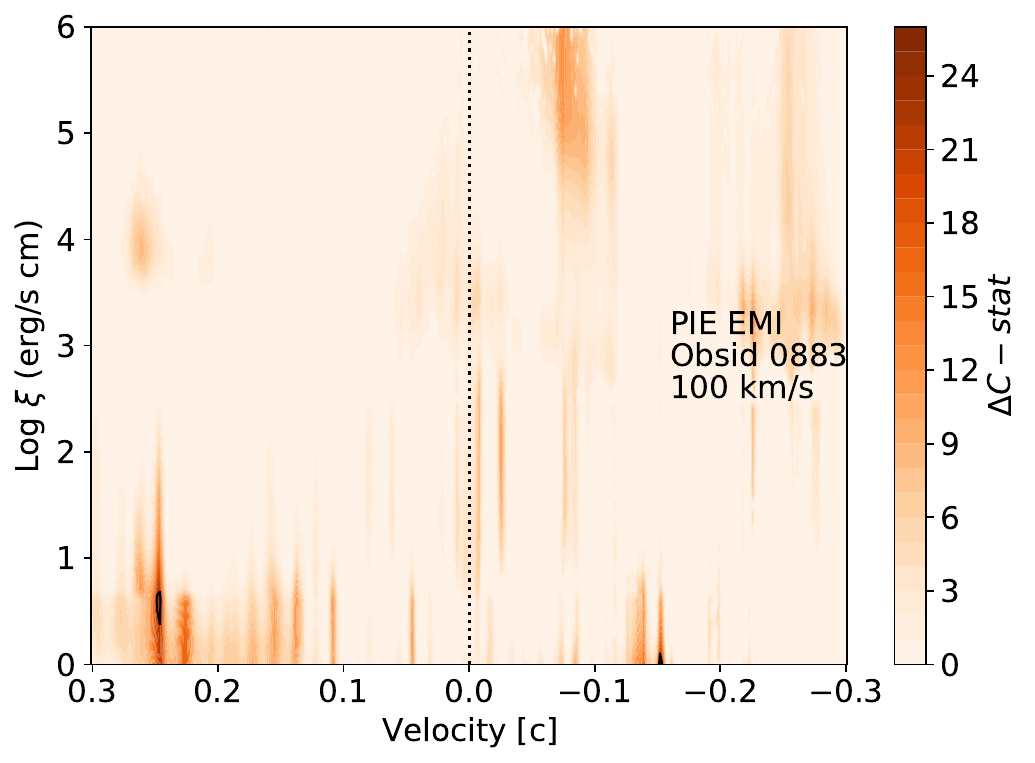}
   \includegraphics[scale=0.325]{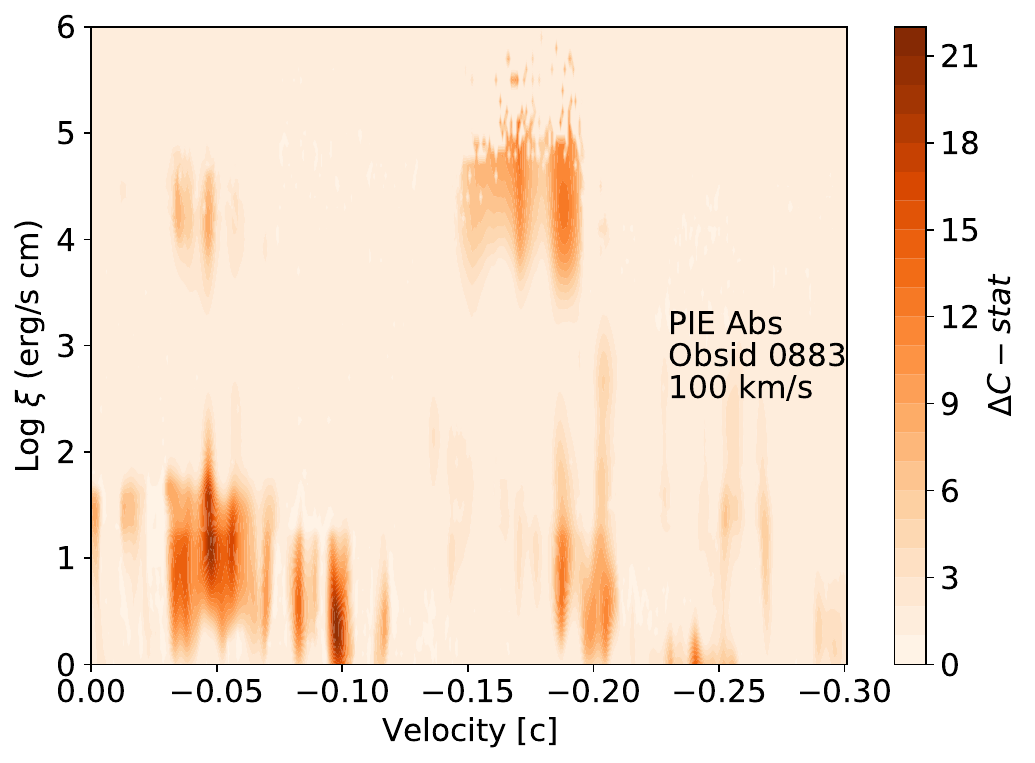}
   \captionof{figure}{{Physical model grids of the individual observations: CIE emission (left), PIE emission (middle) and PIE absorption (right).  The black contours refer to significance levels from 3 to $5\,\sigma$ \textcolor{black}{with steps of $0.5\,\sigma$ estimated with MC} simulations (see Sect \ref{sec:monte_carlo_simulations}).}}
              \label{fig:individual_scans}
    \end{figure*}

{In Fig.\,\ref{fig:individual_scans} we report the multi-dimensional grids of the spectra with different physical models (emission by collisional or photoionised plasmas and absorption by a photoionised plasma) with the same routine used in Sect.\,\ref{sec:model_scan}. The black contours refer to the confidence level estimated through Monte Carlo simulations as described in Sect.\,\ref{sec:monte_carlo_simulations}.}

\begin{figure*}
   \centering
   \includegraphics[scale=0.32]{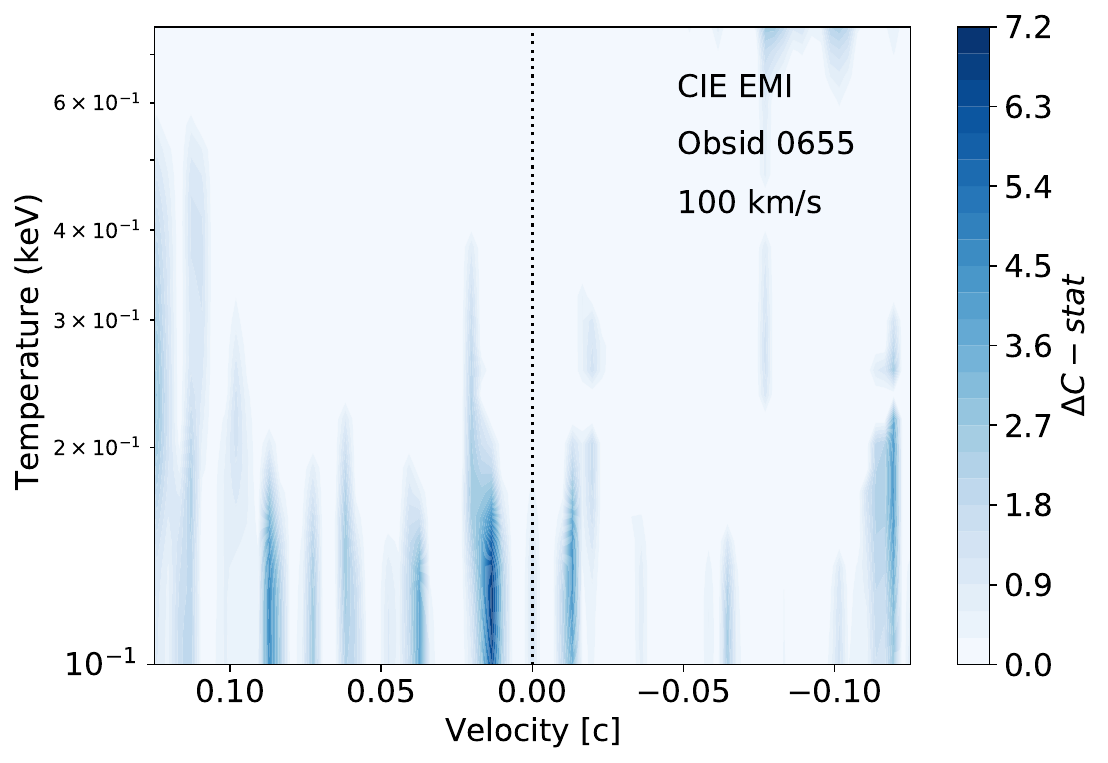}
   \includegraphics[scale=0.32]{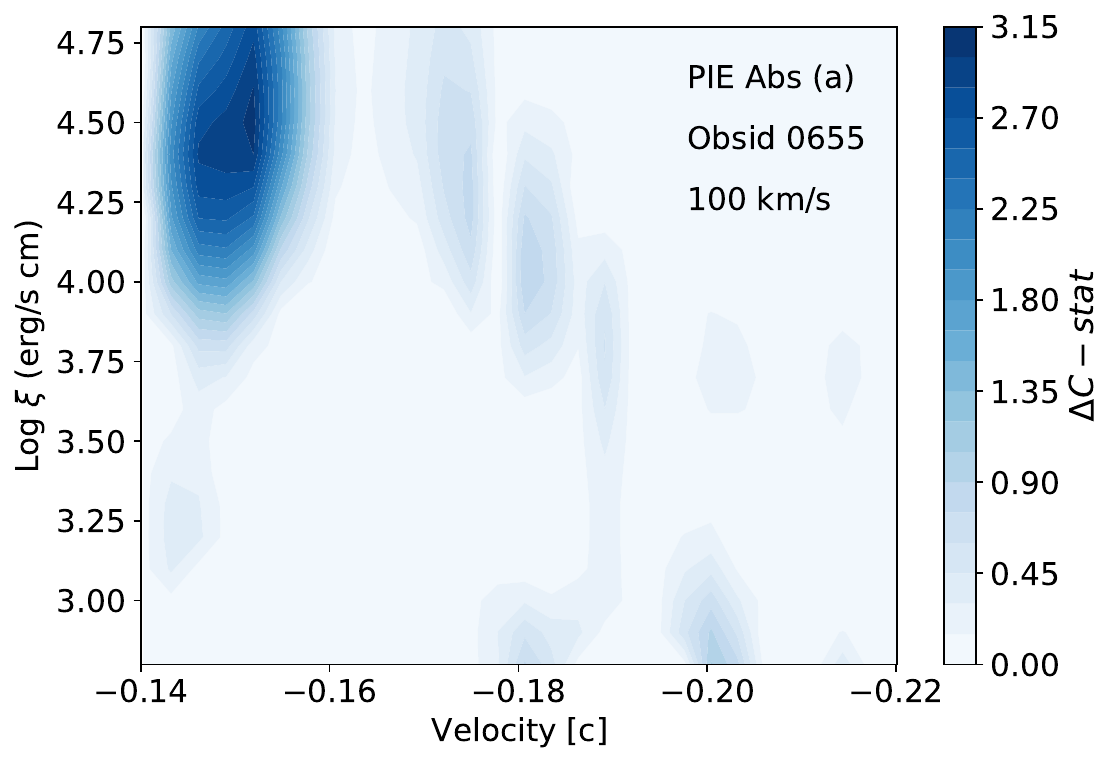}
   \includegraphics[scale=0.32]{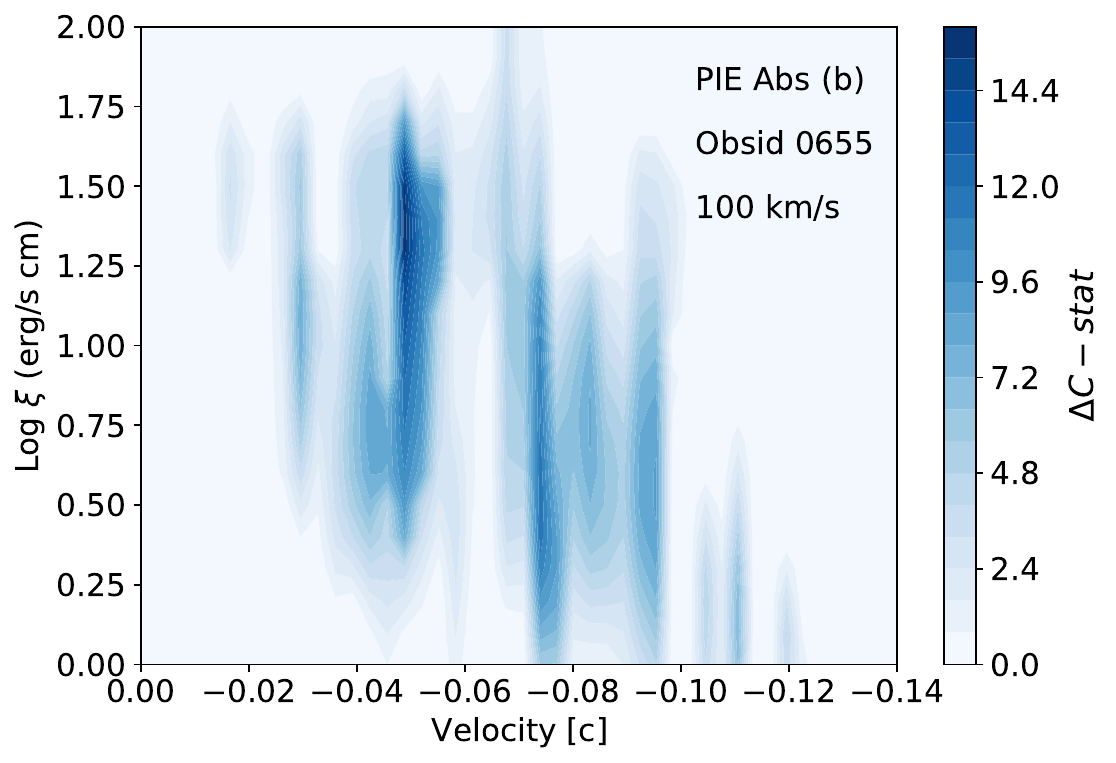}
   \captionof{figure}{{Physical model grids (zoom) of obsid 0655050101 to the address the significance of secondary solutions (therefore the blue color map): CIE emission (left), PIE absorption (part a, middle, and part b, right). Note the much lower $\Delta C$ once each primary solution was taken into account.}}
              \label{fig:secondary_scans}
   \end{figure*}

{In Fig.\,\ref{fig:secondary_scans}, we show further increments when \texttt{cie} and \texttt{xabs} grids were run again for obsid 0655050101 once the contribution of the primary solutions (i.e. those reported in Table\,\ref{table:wind_properties} and Fig.\ref{fig:results_summary_xabs}-\ref{fig:results_summary_cie}) were included in the underlying baseline model. The very low $\Delta C$ show how the primary solution was able to describe most of the corresponding features found with the scans. Moreover, the presence of a low velocity ($-0.05c$) absorption component although rather weak indicates presence of a multiphase plasma as it was already discussed in P17 (see Fig.\,\ref{fig:secondary_scans}, right panel).}

\subsection{Monte Carlo simulations}
\label{appendix:MC_simulations}

\begin{figure*}
   \centering
   \includegraphics[scale=0.3372475]{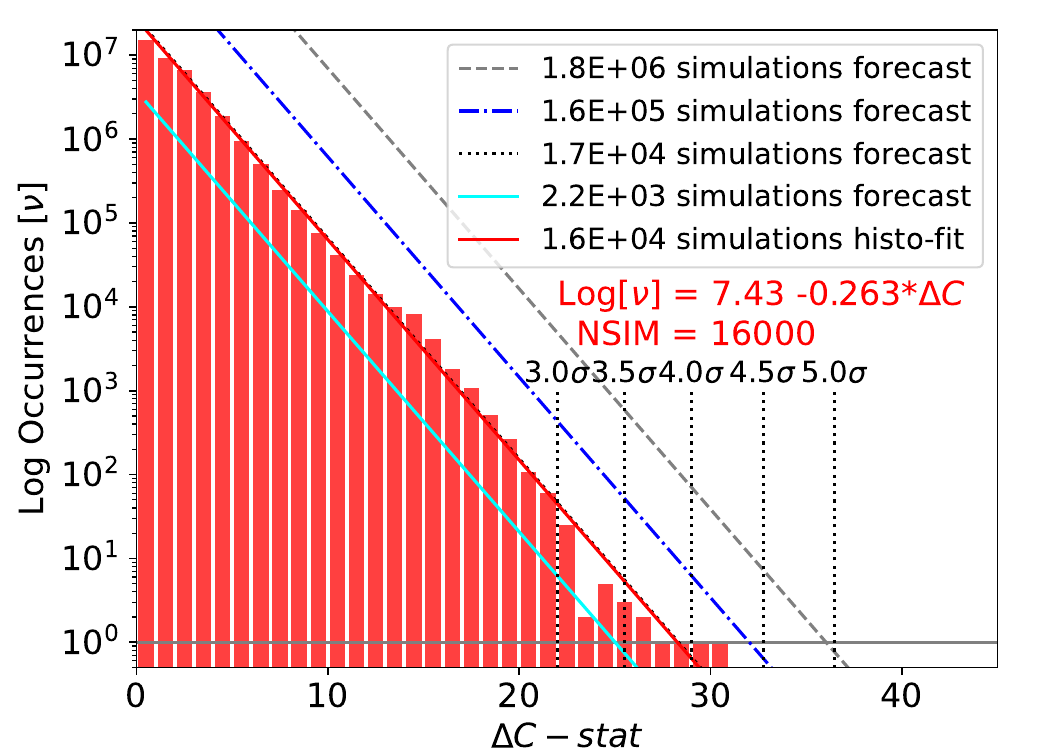}
   \includegraphics[scale=0.3372475]{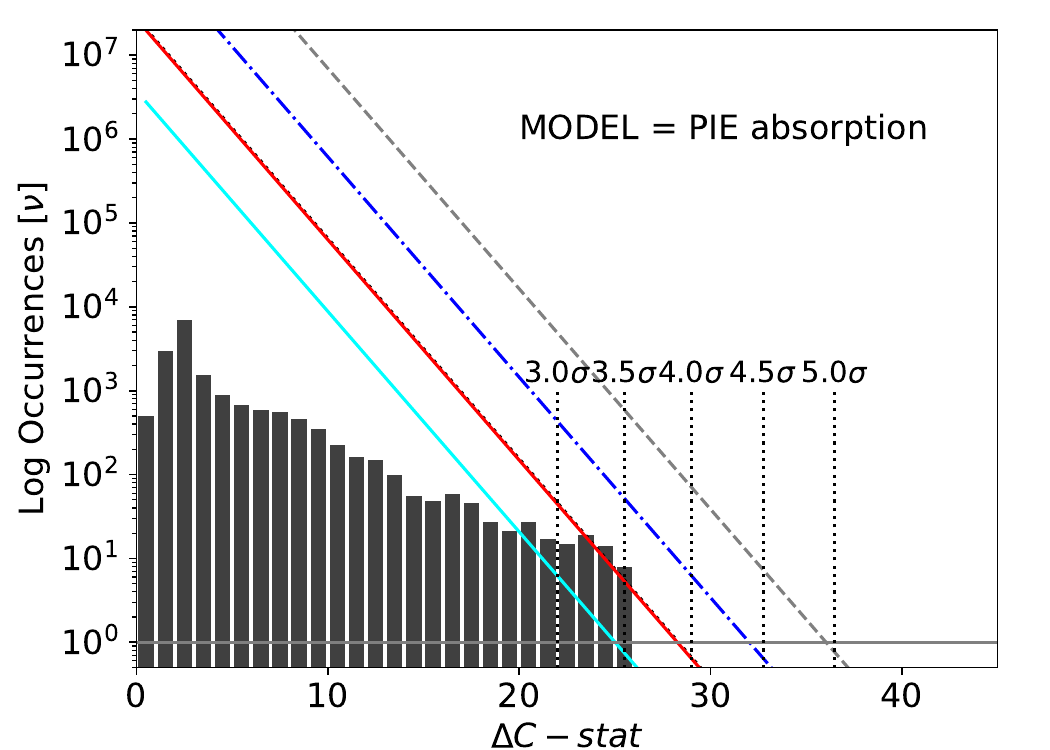}
   \includegraphics[scale=0.3372475]{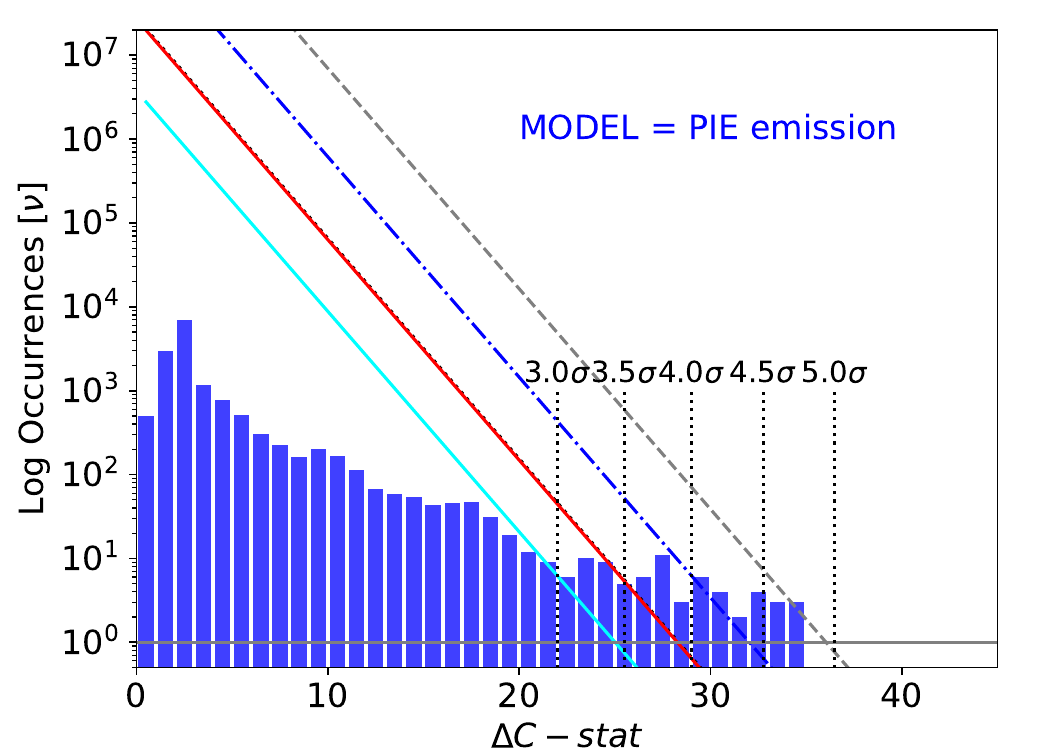}
   \caption{{Monte Carlo simulations for significance estimates (left panel, red bars, template model: continuum model for obsid 0655050101). Forecasts for more than 16,000 simulations are performed adopting a constant histogram slope. As a comparison the results for the \texttt{xabs} (middle panel, black bars) and the \texttt{pion} (right panel, blue bars) histograms for the real data are also shown.}
   }
    \label{fig:mc_simulations}
\end{figure*}

{In Fig.\,\ref{fig:mc_simulations}, we reported with red bars the histogram of the $\Delta C$-stat occurrences obtained from a deep grid of absorption models of photoionised plasma for 16,000 simulated spectra based on the continuum model of RGS+EPIC obsid 0655050101. The stability of the histogram slope ($-0.26$) above about 5,000 simulated spectra enabled us to forecast the results for $1.5\times10^{5}$ and $1.7\times10^{6}$ simulations and probe the $\Delta C$-stat corresponding to detections with confidence levels up to 4.5 and 5.0\,$\sigma$ (see vertical dotted lines). The confidence levels can also be found as black contours in Fig.\,\ref{fig:individual_scans}. The black and blue bars refer to the results obtained for grids of emission and absorption models applied on the real data for the same obsid. For more detail, see Sect.\,\ref{sec:monte_carlo_simulations}.}

\subsection{Contribution of spectral components}
\label{appendix:spectral_components}

\begin{figure*}
   \centering
   \includegraphics[scale=0.685]{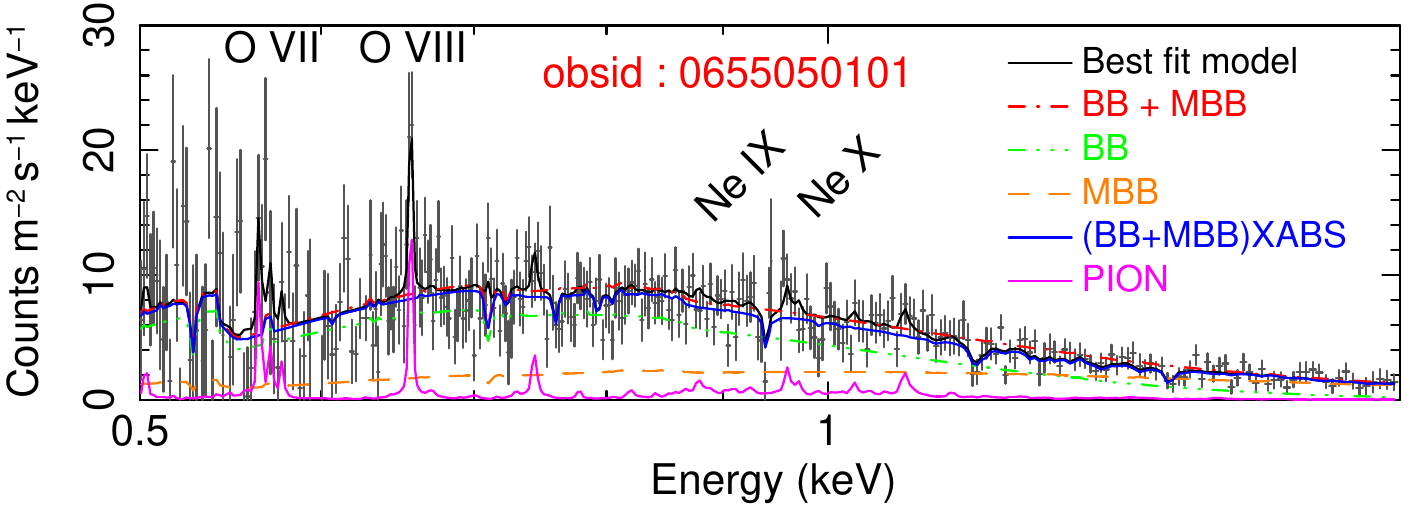}
   \includegraphics[scale=0.685]{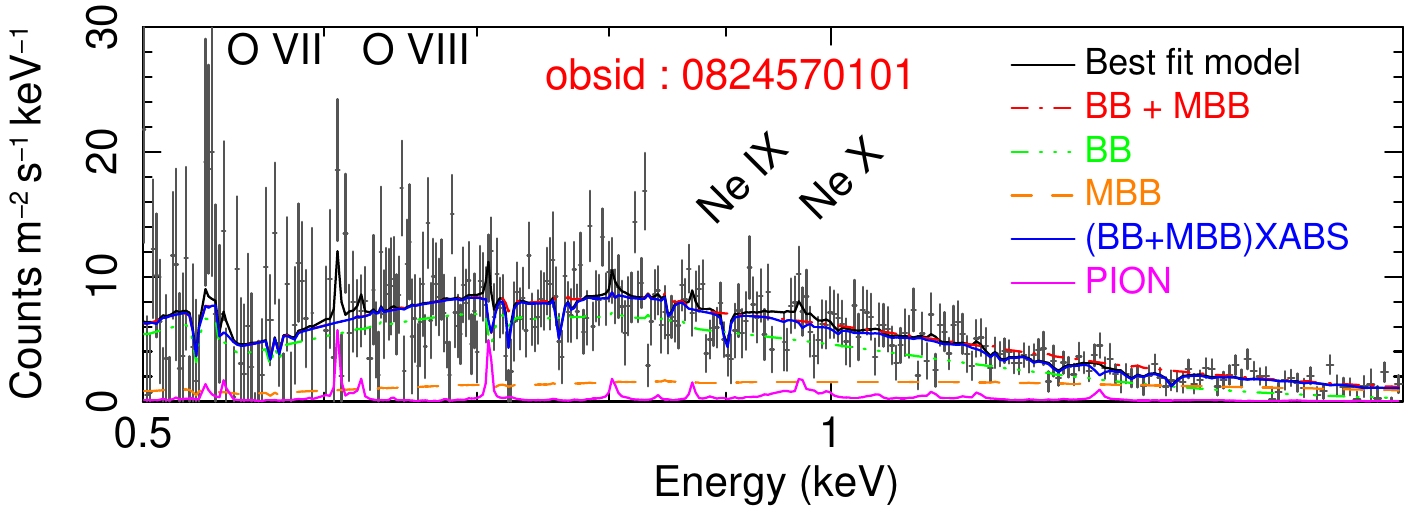}
   \includegraphics[scale=0.685]{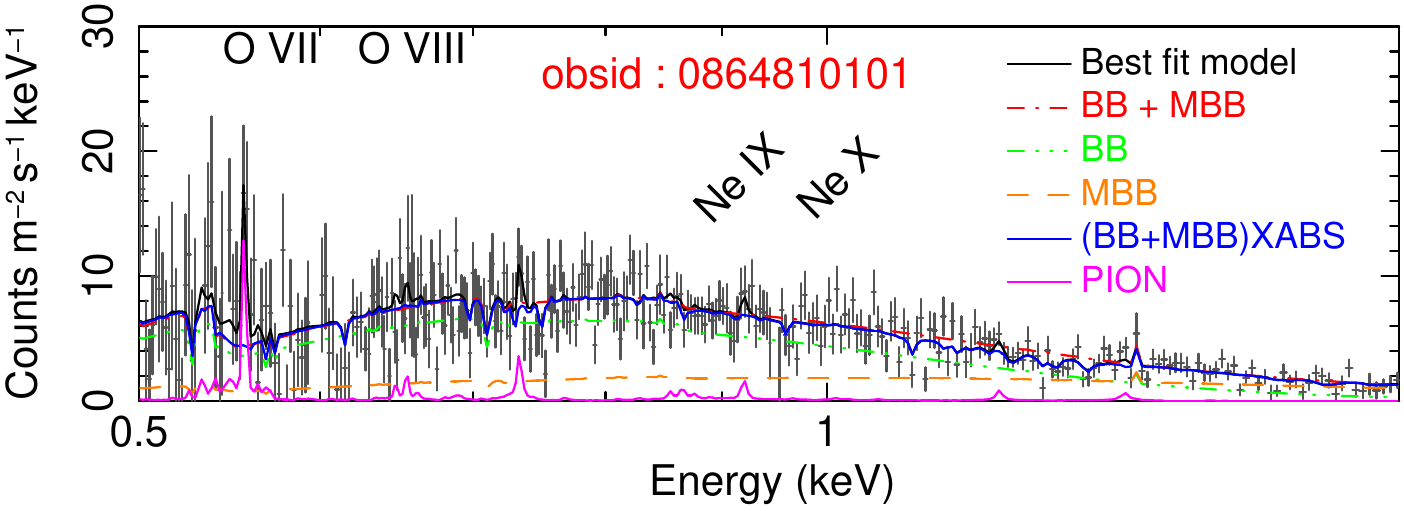}
   \includegraphics[scale=0.685]{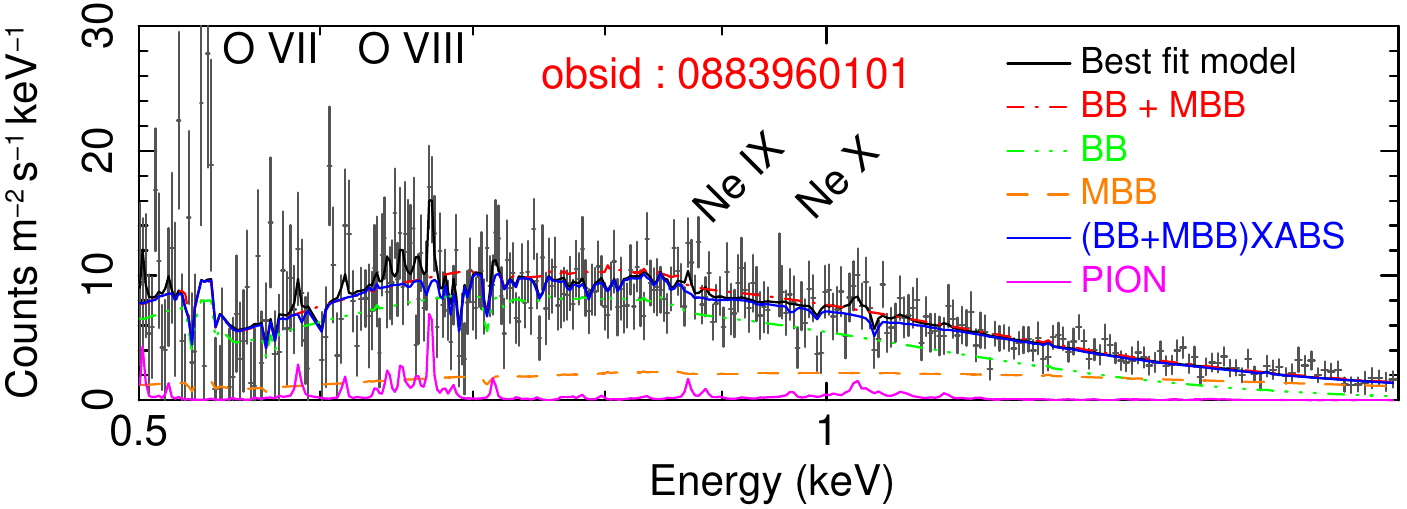}
\vspace{-0.1cm}
   \caption{{Detail on the main spectral components of the continuum (\texttt{bb} and \texttt{mbb}) and the plasma (\texttt{xabs} in absorption and \texttt{pion} in emission) zoomed on the RGS spectra. The best-fit model of each spectrum is also shown in Table\,\ref{table:wind_properties} and previously in Fig.\,\ref{fig:spec_bestfit}. The ISM contribution is present in each model. Labelled are the main ionic species for the soft X-ray emission in their rest-frame (see also Fig.\,\ref{fig:spec_tavg}, \ref{fig:spec_all_gaus} and \ref{fig:spec_all_gaus_lines}).}
   }
    \label{fig:spectral_components}%
\vspace{-0.1cm}
\end{figure*}

{In Fig.\,\ref{fig:spectral_components}, we show in detail the contribution of each spectral component in the best fit for each individual observation. To highlight the contribution of the \texttt{xabs} component we show the case where the simple continuum model was multiplied by the \texttt{xabs} (along with the ISM), i.e. the emission of the \texttt{pion} was removed.}

\end{document}